\documentclass[12pt]{article}
\usepackage{amssymb}
\usepackage{graphicx}
\begin{document}
\newcommand{\beq}{\begin{equation}}
\newcommand{\eeq}{\end{equation}}
\newcommand{\beqa}{\begin{eqnarray}}
\newcommand{\eeqa}{\end{eqnarray}}
\newcommand{\beqar}{\begin{eqnarray*}}
\newcommand{\eeqar}{\end{eqnarray*}}
\newcommand{\al}{\alpha}
\newcommand{\be}{\beta}
\newcommand{\del}{\delta}
\newcommand{\D}{\Delta}
\newcommand{\eps}{\epsilon}
\newcommand{\ga}{\gamma}
\newcommand{\Ga}{\Gamma}
\newcommand{\ka}{\kappa}
\newcommand{\nn}{\nonumber}
\newcommand{\inn}{\!\cdot\!}
\newcommand{\h}{\eta}
\newcommand{\ii}{\iota}
\newcommand{\kk}{\varphi}
\newcommand\F{{}_3F_2}
\newcommand{\la}{\lambda}
\newcommand{\La}{\Lambda}
\newcommand{\na}{\prt}
\newcommand{\Om}{\Omega}
\newcommand{\om}{\omega}
\newcommand{\p}{\phi}
\newcommand{\sig}{\sigma}
\renewcommand{\t}{\theta}
\newcommand{\z}{\zeta}
\newcommand{\ssc}{\scriptscriptstyle}
\newcommand{\eg}{{\it e.g.,}\ }
\newcommand{\ie}{{\it i.e.,}\ }
\newcommand{\labell}[1]{\label{#1}} 
\newcommand{\reef}[1]{(\ref{#1})}
\newcommand\prt{\partial}
\newcommand\veps{\varepsilon}
\newcommand{\pol}{\varepsilon}
\newcommand\vp{\varphi}
\newcommand\ls{\ell_s}
\newcommand\cF{{\cal F}}
\newcommand\cA{{\cal A}}
\newcommand\cS{{\cal S}}
\newcommand\cT{{\cal T}}
\newcommand\cV{{\cal V}}
\newcommand\cL{{\cal L}}
\newcommand\cM{{\cal M}}
\newcommand\cN{{\cal N}}
\newcommand\cG{{\cal G}}
\newcommand\cH{{\cal H}}
\newcommand\cI{{\cal I}}
\newcommand\cJ{{\cal J}}
\newcommand\cl{{\iota}}
\newcommand\cP{{\cal P}}
\newcommand\cQ{{\cal Q}}
\newcommand\cg{{\it g}}
\newcommand\cR{{\cal R}}
\newcommand\cB{{\cal B}}
\newcommand\cO{{\cal O}}
\newcommand\tcO{{\tilde {{\cal O}}}}
\newcommand\bz{\bar{z}}
\newcommand\bc{\bar{c}}
\newcommand\bw{\bar{w}}
\newcommand\bX{\bar{X}}
\newcommand\bK{\bar{K}}
\newcommand\bA{\bar{A}}
\newcommand\bZ{\bar{Z}}
\newcommand\bxi{\bar{\xi}}
\newcommand\bphi{\bar{\phi}}
\newcommand\bpsi{\bar{\psi}}
\newcommand\bprt{\bar{\prt}}
\newcommand\bet{\bar{\eta}}
\newcommand\btau{\bar{\tau}}
\newcommand\hF{\hat{F}}
\newcommand\hA{\hat{A}}
\newcommand\hT{\hat{T}}
\newcommand\htau{\hat{\tau}}
\newcommand\hD{\hat{D}}
\newcommand\hf{\hat{f}}
\newcommand\hg{\hat{g}}
\newcommand\hp{\hat{\phi}}
\newcommand\hi{\hat{i}}
\newcommand\ha{\hat{a}}
\newcommand\hb{\hat{b}}
\newcommand\hQ{\hat{Q}}
\newcommand\hP{\hat{\Phi}}
\newcommand\hS{\hat{S}}
\newcommand\hX{\hat{X}}
\newcommand\tL{\tilde{\cal L}}
\newcommand\hL{\hat{\cal L}}
\newcommand\tG{{\tilde G}}
\newcommand\tg{{\tilde g}}
\newcommand\tphi{{\widetilde \phi}}
\newcommand\tPhi{{\widetilde \Phi}}
\newcommand\te{{\tilde e}}
\newcommand\tk{{\tilde k}}
\newcommand\tf{{\tilde f}}
\newcommand\ta{{\tilde a}}
\newcommand\tb{{\tilde b}}
\newcommand\tc{{\tilde c}}
\newcommand\td{{\tilde d}}
 \newcommand\tR{{\tilde R}}
\newcommand\teta{{\tilde \eta}}
\newcommand\tF{{\widetilde F}}
\newcommand\tK{{\widetilde K}}
\newcommand\tE{{\widetilde E}}
\newcommand\tpsi{{\tilde \psi}}
\newcommand\tX{{\widetilde X}}
\newcommand\tD{{\widetilde D}}
\newcommand\tO{{\widetilde O}}
\newcommand\tS{{\tilde S}}
\newcommand\tB{{\widetilde B}}
\newcommand\tA{{\widetilde A}}
\newcommand\tT{{\widetilde T}}
\newcommand\tC{{\widetilde C}}
\newcommand\tV{{\widetilde V}}
\newcommand\thF{{\widetilde {\hat {F}}}}
\newcommand\Tr{{\rm Tr}}
\newcommand\tr{{\rm tr}}
\newcommand\STr{{\rm STr}}
\newcommand\hR{\hat{R}}
\newcommand\M[2]{M^{#1}{}_{#2}}

\newcommand\bS{\textbf{ S}}
\newcommand\bI{\textbf{ I}}
\newcommand\bJ{\textbf{ J}}

\begin{titlepage}
\begin{center}

\vskip 2 cm
{\LARGE \bf   $\alpha'$-corrections to DBI action  \\  \vskip 0.25 cm 
via T-duality constraint
 }\\
\vskip 1.25 cm
  Saman Karimi\footnote{karimi.saman@mail.um.ac.ir} and  Mohammad R. Garousi\footnote{garousi@um.ac.ir}

\vskip 1 cm
{{\it Department of Physics, Faculty of Science, Ferdowsi University of Mashhad\\}{\it P.O. Box 1436, Mashhad, Iran}\\}
\vskip .1 cm
 \end{center}

\begin{abstract}

It is known that D$_p$-brane effective action at the leading order of $\alpha'$ in flat space-time which is given by DBI action, transforms to D$_{p-1}$-brane effective action under standard T-duality transformations of the open string  gauge bosons and transverse scalar fields.  Extending this duality to     order   $\alpha'$, one may find corrections to the DBI action which include the second fundamental form $\Omega$ and the covariant derivative of gauge field strength $DF$, as well as  the corrections to the   T-duality transformations. Using this idea, up to two parameters,  we have found all 81   covariant couplings of   $D FD F$  and $\Omega\Omega$   with zero, two, four and six   $F$'s.   The four gauge field couplings that the T-duality constraint fixes are consistent with  the known couplings in the literature. 
\end{abstract}

\end{titlepage}

\section{Introduction}

One of the most exciting discoveries in  string theory is   T-duality \cite{Giveon:1994fu,Alvarez:1994dn}. This   duality   may be used to construct the effective field theory at low energy which may provide a manifestly background independent formulation of string theory \cite{Hohm:2016lge,Hohm:2018zer}. One approach for constructing this effective action  is the   Double Field Theory      \cite{Siegel:1993xq,Siegel:1993th,Siegel:1993bj,Hull:2009mi,Aldazabal:2013sca}  in which the   T-duality is manifest,    as the effective action is  $O(D,D)$-invariant by constructions. However, coordinate transformations in this approach recive $\alpha'$ corrections \cite{Hohm:2014xsa,Marques:2015vua}. 
Another  T-duality based approach for constructing the   effective action, is to use the constraint that   the dimensional reduction of an effective action   on a circle  must be invariant under the T-duality transformations \cite{Garousi:2017fbe}. In this   approach, the couplings are invariant under the standard coordinate transformations, however, the T-duality transformations   receive $\alpha'$-corrections   \cite{Bergshoeff:1995cg,Kaloper:1997ux}.    Using the T-duality constraint, the standard gravity and dilaton  couplings in the effective actions at orders $\alpha',\alpha'^2,\alpha'^3$ have been reproduced  in \cite{Razaghian:2017okr,Razaghian:2018svg}. It has been  observed in \cite{Razaghian:2017okr} that the form of $\alpha'$ corrections to the Buscher rules depends on the scheme that one uses for the effective action. 

The effective field theory of a $ {D_p} $-brane in bosonic string theory includes various world-volume couplings of open string tachyon, transverse scalar fields, closed string tachyon, graviton, dilaton and B-field. Because of tachyons, the bosonic string theory and its $ {D_p} $-branes are all unstable. Assuming the tachyons are frozen at the top of their corresponding tachyon potentials, the effective action at the leading order of $\alpha'$ in flat spacetime is then given by DBI action \cite{Leigh:1989jq,Bachas:1995kx}:
\begin{eqnarray}
{S_p} \supset  - {T_p}\int {{d^{p + 1}}\sigma \sqrt { - \det ({{\tilde G}_{ab}} + {F_{ab}})} } \labell{a1} 
\end{eqnarray}
where $ {T_p} $ is tension of $ {D_p} $-brane,   $ {{F_{ab}}} $ is gauge field strength of $ {A_a} $  and $ {{{\tilde G}_{ab}}} $ is metric which is pull-back of the bulk flat metric onto the world-volume \footnote{Our index convention is that the Greek letters $ (\mu ,\nu ,...) $ are the indices of the space-time coordinates, the Latin letters, $ (a,b,c,...) $ are the world-volume indices and the letters $ (i,j,k,...) $ are the normal bundle indices. The killing coordinate   $y$ is along the  world-volume. The world-volume indices after the reduction of D$_p$-brane to D$_{p-1}$-brane  are $ (\tilde a,\tilde b,\tilde c,...). $
}\ie
\begin{eqnarray}
  {{\tilde G}_{ab}} = P{[\eta ]_{ab}} &=& \frac{{\partial {X^\mu }}}{{\partial {\sigma ^a}}}\frac{{\partial {X^\nu }}}{{\partial {\sigma ^b}}}{\eta _{\mu \nu }} \hfill\nonumber \\
   &=& {\eta _{ab}} + {\partial _a}{\chi ^i}{\partial _b}{\chi ^j}{\eta _{ij}} \label{a2} 
\end{eqnarray}
where $ {{X^\mu }} $ is coordinate of space-time and $ {\eta _{\mu \nu }} $ is flat space-time metric. In the second line the pull-back is written in the static guage, \ie  $ {X^a} = {\sigma ^a} $ and  $ {X^i} = {\chi ^i} $.
The DBI action \reef{a1} is invariant  under the general coordinate transformations and is covariant  under the standard T-duality transformation \cite{Myers:1999ps}. With our normalization for the gauge field, the DBI action is at the leading order of $\alpha'$. It involves infinite number of $F$ and ${\partial  }{\chi ^i}{\partial }{\chi ^j}{\eta _{ij}}$.
  The first correction to this action is at order  $\alpha'$ which includes $DFDF$ or $\Omega\Omega$ and infinite number of   $F$'s. 
The
higher derivative 
corrections to the Born-Infeld action in the bosonic and superstring theories, for only gauge field, have been studied in
\cite{Abouelsaood:1986gd,Tseytlin:1987ww,Andreev:1988cb, Wyllard:2000qe, Andreev:2001xx, Wyllard:2001ye}.

The world-volume couplings in the DBI action in the string frame are independent of $p$, the dimension of the $D_p$-brane. This has been used in \cite{Myers:1999ps} to observes that the DBI action is  covariant under T-duality transformation.  Assuming the higher derivative couplings on the world-volume of $D_p$-brane are also independent of the dimension of the brane, one expects the effective action of $D_p$-brane at any order of $\alpha'$   to be covariant under the T-duality transformation.  Using this constraint, we are going to study the $\alpha'$ corrections to the DBI action in this paper. Since there are infinite number of   $F$'s involved in the couplings at order $\alpha'$, we consider couplings which have zero, two, four and six   $F$'s. The couplings which have zero  $F$ are  
\begin{eqnarray}
{S_p} &\supset&  - {\alpha'}{T_p}\int {{d^{p + 1}}\sigma \sqrt { - \det ({{\tilde G}_{ab}} )} } \Big[C  {{\tilde  \bot }_{\mu \nu }}{{\tilde G}^{ab}}{{\tilde G}^{cd}}(\Omega _{ab}^{\,\,\,\,\,\mu }\Omega _{cd}^{\,\,\,\,\,\nu } - \Omega _{ac}^{\,\,\,\,\,\mu }\Omega _{bd}^{\,\,\,\,\,\nu })
\Big]\labell{a6}
\end{eqnarray}
where $C$ is a constant, $\tG^{ab}$ is inverse of the pull-back metric and the second fundamental form $ \Omega  $  in the bosonic theory  is defined to be \cite{Corley:2001hg}\footnote{The second fundamental form in the superstring theory is defined in \cite{Bachas:1999um} to be 
\beqa
  \Omega _{ab}^{\,\,\,\,\,\alpha} &=&\frac{\prt^2 X^{\alpha}}{\prt\sigma^a\prt\sigma^b}-\frac{\prt X^\alpha}{\prt\sigma_c}{\tilde\Gamma}_{ab}{}^c+\frac{\prt X^{\mu}}{\prt\sigma^a}\frac{\prt X^{\nu}}{\prt\sigma^b}\Gamma_{\mu\nu}{}^{\alpha}\nn
\eeqa
where ${\tilde\Gamma}_{ab}{}^c$ is the connection made of the pull-back metric. In flat spacetime  it becomes
\beqa
\Omega _{ab}{}^{\alpha} &=& \tilde  \bot^{\alpha\beta}\eta_{\beta\mu}{\partial _a}{\partial _b}{X ^\mu}
\eeqa
where we have used the definition $\tG_{ab}=\prt_a X^\mu\prt_b X^\nu \eta_{\mu\nu}$. Using the fact that in the effective action, the spacetime index of the second fundamental form is always contracted with the projection tensor $\tilde\bot_{\mu\nu}$, the projection tensor $\tilde\bot^{\alpha\beta}$ in the above expression can be removed.  Then the second fundamental form becomes $\Omega _{ab}{}^{\alpha} ={\partial _a}{\partial _b}{X ^\alpha}$ which is the same as \reef{a5} in the static gauge.}
\begin{eqnarray}
  \Omega _{ab}^{\,\,\,\,\,\alpha} &=&\frac{\prt^2 X^{\alpha}}{\prt\sigma^a\prt\sigma^b}+\frac{\prt X^{\mu}}{\prt\sigma^a}\frac{\prt X^{\nu}}{\prt\sigma^b}\Gamma_{\mu\nu}{}^{\alpha}\labell{aaa}
\end{eqnarray}
  The tensor $ {{\tilde  \bot }_{\mu \nu }}$ in \reef{a6} is a projection operator, \ie $ {\eta ^{\nu \alpha }}{{\tilde  \bot }_{\mu \nu }}{{\tilde  \bot }_{\alpha \beta }} = {{\tilde  \bot }_{\mu \beta }} $, which projects space-time tensors to the transverse space. It is defined as    ${{\tilde  \bot }_{\mu \nu }} = {\eta _{\mu \nu }} - \eta_{\mu\alpha}\eta_{\nu\beta}{{\tilde G}^{\alpha \beta }} $ where the  
   first fundamental form  $ {{\tilde G}^{\mu \nu }} $ is defined as 
\begin{eqnarray}
{{\tilde G}^{\mu \nu }} = \frac{{\partial {X^\mu }}}{{\partial {\sigma ^a}}}\frac{{\partial {X^\nu }}}{{\partial {\sigma ^b}}}{{\tilde G}^{ab}} \label{a4}
\end{eqnarray}
which is another  projection operator, \ie $ {\eta _{\nu \alpha }}{{\tilde G}^{\mu \nu }}{{\tilde G}^{\alpha \beta }} = {{\tilde G}^{\mu \beta }} $. It projects space-time tensors to the world-volume.  

In flat spacetime and in static gauge, the second fundamental form \reef{aaa} is zero when the spacetime index $\alpha$ is a world volume and it is the second derivative of the transverse scalar fields when $\alpha$ is a transverse index, \ie
\beqa
 \Omega _{ab}^{\,\,\,\,\,\alpha} &=&{\partial _a}{\partial _b}{\chi ^i}\delta_{i}{}^{\alpha} \labell{a5}
\eeqa
The covariant action \reef{a6} includes infinite number of transverse scalar fields through the expansion of pull-back metric.   We have chosen the relative coefficients of the two terms in  \reef{a6} to have no corrections to the propagators of   the transverse scalar fields. This action, however, is not total derivative term for terms with more tham two transverse scalars. The coefficient $C$ is a  parameter which should be fixed by some calculations in string theory, \eg by studying the S-matrix element of two gravitons off the D$_p$-brane this parameter has been found in \cite{Corley:2001hg} to be $C=1$  . 
    There are similar actions  with some extra $F$'s, which we will find some of them in section 2. The parameters in these  couplings and in \reef{a6} may be found by S-matrix calculations, however, we are interested in this paper to find them by imposing the T-duality constraint. 

There are also couplings at order $\alpha'$ which include $DFDF$ and some extra $F$'s. The covariant derivative of $F$ is 
\begin{eqnarray}
  {D_a}{F_{bc}} &=& {\partial _a}{F_{bc}} - \tilde \Gamma _{ab}^{\,\,\,\,\,d}{F_{dc}} - \tilde \Gamma _{ac}^{\,\,\,\,\,d}{F_{bd}} \hfill \nonumber\\
   &=& {\partial _a}{F_{bc}} - \eta_{ij}\tilde G^{de}\prt_e\chi^i{\partial _a}{\partial _b}{\chi ^j}{F_{dc}} + \eta_{ij}\tilde G^{de}\prt_e\chi^i{\partial _a}{\partial _c}{\chi ^j}{F_{db}} \labell{a3}
\end{eqnarray}
where  the christoffel symbol $ \tilde \Gamma _{ab}^{\,\,\,\,\,c} $  is made of the pull-back metric   $ {{\tilde G}_{ab}} $.
As we will see in the next section, at the level of zero extra $F$, the couplings are total derivative terms, and at the level of two and more extra $F$'s, there are nontrivial couplings that their coefficients may be found by the T-duality constraint.   As we will see, all parameters in the actions with zero, two, four and six  extra $F$'s for which  we have done the calculations  explicitly, can be fixed up to two parameters. We choose one of them to be the coefficient $C$ which is fixed by the S-matrix calculations to be $C=1$.

The outline of the paper is as follows: In section 2, we find   all independent couplings of $DFDF$ and  $\Omega\Omega$ with two, four and six extra $F$'s. To this end, we first write all contractions  of $DFDF$ and  $\Omega\Omega$ with two, four and six  $F$'s. The terms involving  $\Omega\Omega$ are all independent, however, the terms involving  $DFDF$ are not all independent as they are related by total derivative terms and the Bianchi identity. We introduce a method for imposing the Bianchi identity to find all independent couplings. In section 3, we impose the T-duality constraint on the independent couplings found in section 2 to fix their corresponding unknown coefficients in terms of two parameters. We show that the coefficients of the four gauge field couplings that the T-duality constraint fixes are consistent with  the coefficients that one finds by the S-matrix method. We   find also covariant  couplings of six and eight gauge fields which have not been found   by the S-matrix method.

\section{Independent couplings}

In this section we are going to find $DFDF$ and $\Omega\Omega$ couplings with two, four and six extra $F$'s. We begin with the couplings with two extra $F$'s. There are 18 contractions with structure $FFDFDF$. However, not all of them are independent\footnote{We use the mathematica package 'xAct' \cite{Nutma:2013zea} for performing the calculations in this paper.} . Some of them are related by total derivative terms and some other terms are related by the Bianchi identity $D_{[a}F_{bc]}=0$. Note that using integration by part one can easily observe that the couplings with structure $FFFDDF$ can be written in terms of $FFDFDF$. To find the independent couplings we first construct the current $I^a$ from 9 contractions of terms with structure $FFFDF$. The 9 total derivative terms $D[FFFDF]$, however, produce   terms with structures $FFFDDF$ and $FFDFDF$. The two covariant derivatives in $D_aD_b F_{cd}$   can be written as symmetric and antisymmetric parts, \ie 
\beqa
D_aD_b F_{cd}&=&\frac{1}{2}\{D_a,D_b\}F_{cd}+\frac{1}{2}[D_a,D_b]F_{cd}\labell{anti}  
\eeqa
The antisymmetric part is identical to $\tR F$. On the other hand, using the Gauss-Codazzi equation 
\beqa
\tR_{abcd}&=&{\tilde  \bot }_{ij}(\Omega_{ac}{}^i\Omega_{bd}{}^j-\Omega_{ad}{}^i\Omega_{bc}{}^j)
\eeqa
the antisymmetric part in \reef{anti} produces couplings with structure $FFFF\Omega\Omega$. They will change the unknown coefficients in the couplings with structure $FFFF\Omega\Omega$. Hence, if one   uses all contractions of    $FFFF\Omega\Omega$, with arbitrary coefficients, as independent couplings, one is allowed to ignore the antisymmetric part in  $D_aD_b F_{cd}$, \ie the two covariant derivatives  is symmetric  . Using this symmetry, one finds there are 6 terms in total derivative terms which have structure $FFFDDF$. Constraining them to be zero, one finds 3 total derivative terms with structure $FFDFDF$. Adding these terms to the 18 contractions with structure $FFDFDF$, one can reduce them to 15 terms by choosing the coefficients of the total derivative terms to eliminate 3 terms. We choose to eliminate the 3 terms which do note include $D_aF^{ab}$, because as we will discuss in a moment they can be eliminated by field redefinitions. 

Now one has to impose the Bianchi identity on $DF$-terms  as well.   Writing   the first term on the right hand side of \reef{a3} in terms of potential  $F_{ab}=\prt_aA_b-\prt_bA_a$ one can write the covariant derivative of $F$   as $D_aF_{bc}=F'_{abc}-F'_{acb}$ where the function $F'_{abc}$ which is not gauge invariant,  is symmetric with respect to its first two indices. 
      Writing $DF$ in terms of $F'$, one can easily observe that the left hand side of the Bianchi identity, \ie $D_{[a}F_{bc]}=0$,  is zero. 

When one rewrites the 15 couplings in terms of  $F'$ , one would find 7 independent couplings. Therefore, the Bianchi identity reduces the 15 couplings to 7 independent couplings when they are written in terms   $F'_{abc}$. There are different ways to write the 7 independent couplings in terms of field strength $F_{ab}$. One particular choice for the couplings is   
\begin{eqnarray}
  {F_{de}}{F^{de}}{D_a}{F_{bc}}{D^a}{F^{bc}}&,&\,\,\,\,\,\,F_c^e{F_{de}}{D^a}{F^{bc}}{D_a}F_b^d \hfill \nonumber\\
  F_a^e{F_{de}}{D^a}{F^{bc}}{D^d}{F_{bc}}&,&\,\,\,\,\,\,F_c^e{F_{de}}{D^a}F_a^c{D^b}F_b^d \hfill \nonumber\\
  {F_{cd}}{F_{be}}{D^a}F_a^c{D^b}{F^{de}}&,&\,\,\,\,\,\,{F_{de}}{F^{de}}{D^a}F_a^c{D^b}{F_{cb}} \hfill \nonumber\\
  F_b^e{F_{de}}{D^a}F_a^c{D^b}F_c^d \labell{a12}
\end{eqnarray}
where the indices are raised by the inverse metric $\tG^{ab}$. Our notation for $F_a^b$ is that the earlier alphabet index appears first. All other choices for the couplings are identical to the above couplings after using the Bianchi identity, \ie they all are identical when they are written in terms of potential $F'_{abc}$. Similar calculations for $DFDF$ with zero extra $F$ produces no independent coupling.

The last four terms in \reef{a12}  include $D_aF^{ab}$. Under field redefinition $A_a\rightarrow A_a+\delta A_a$, $\chi^i\rightarrow \chi^i+\delta\chi^i$ the DBI action   produces the couplings
\beqa
\sqrt{-\det(\tG)}\Big[\frac{1}{2}D_aF^{ab}\delta A_b+\tG^{ab}\Omega_{ab}{}^j\delta\chi^i\eta_{ij}+\cdots\Big]\labell{fred}
\eeqa
where dots represent terms which have some powers of $F$. Hence, the coefficients of the couplings   which include $D_aF^{ab}$  or $\Omega^a{}_{a}{}^i$ can be changed  under field redefinitions. On the other hand, it has been observed in \cite{Razaghian:2017okr} that the corrections to the T-duality transformations depend on the scheme that one uses for the field variables. For simplicity we use the scheme in which there are minimum number of couplings, \ie we use  the field redefinitions to eliminate all terms which include $D_aF^{ab}$.   So up to field redefinitions, there are 3 independent couplings in \reef{a12}.

There are 5 independent couplings with structure   $ FF\Omega \Omega $, \ie
\begin{eqnarray}
{F_{ab}}{F_{cd}}{\Omega ^{aci}}\Omega _{\,\,\,\,\,\,i}^{bd},\,\,\,F_a^{\,c}{F_{bc}}{\Omega ^{adi}}\Omega _{\,\,\,di}^b,\,\,\,{F_{bc}}{F^{bc}}{\Omega _{adi}}{\Omega ^{adi}},\,\,\,{F_{bc}}{F^{bc}}\Omega _{\,\,\,a}^{a\,\,\,\,i}\Omega _{\,\,\,di}^d,\,\,\,F_c^{\,d}{F_{bd}}\Omega _{\,\,\,a}^{a\,\,\,\,i}\Omega _{\,\,\,\,\,\,i}^{cb} \label{a10}
\end{eqnarray}
where the world-volume indices are raised by $\tG^{ab}$ and the transverse indices are lowered by ${\tilde  \bot }_{ij}$. Using the variation \reef{fred}, one can use a scheme in which the last two terms are eliminated by appropriate field redefinitions\footnote{ One could also use field redefinition to remove the first term in \reef{a6}, however, the absence of this term changes the propagator of the scalar fields. In that case, the $\alpha'$ corrections to the T-duality transformations would have linear term as well as nonlinear terms. We work in this paper with the couplings \reef{a6}. }.   All together, up to field redefinitions there are 6 independent terms at two extra $F$ level, \ie
\begin{eqnarray}
  {S_p} &\supset&  - {\alpha'}{T_p}\int {{d^{p + 1}}\sigma \sqrt { - \det ({{\tilde G}_{ab}}  )} } \Big[ {C_1}{F_{ab}}{F_{cd}}{\Omega ^{aci}}\Omega _{\,\,\,\,\,\,i}^{bd}    + {C_2}F_a^{\,c}{F_{bc}}{\Omega ^{adi}}\Omega _{\,\,\,di}^b \nn\\
   &&\qquad\qquad\qquad\qquad\qquad\qquad + {C_3}{F_{bc}}{F^{bc}}{\Omega _{adi}}{\Omega ^{adi}}   + {N_1}{F_{de}}{F^{de}}{D_a}{F_{bc}}{D^a}{F^{bc}} \nn\\ 
 && \qquad\qquad\qquad\qquad\qquad\qquad + {N_2}F_c^e{F_{de}}{D^a}{F^{bc}}{D_a}F_b^d+ {N_3}F_a^e{F_{de}}{D^a}{F^{bc}}{D^d}{F_{bc}} \Big] \labell{a11}
\end{eqnarray}
The coefficients $C_1,C_2,C_3, N_1,N_2,N_3$  and $C$ in \reef{a6}   are 7 parameters that can be found by the S-matrix elements of four open string vertex operators   \cite{Andreev:1988cb, Garousi:2015qgr}. They are 
\beqa
 C=1\,\,;\,\,C_1=C_2=1\,\,,\,\,C_3=-\frac{1}{4}&;&N_1=-\frac{1}{24}\,\,,\,\, N_2=-\frac{1}{3}\,\,,\,\, N_3=\frac{1}{6}\labell{b4}
\eeqa
However, we are going to find them in the next section  by imposing the T-duality constraint. 

At the level of four extra $F$'s, there are 56 contractions with structure $FFFFDFDF$.  To find the total derivative terms, we note that there are 21 total derivative terms with structure $D[FFFFFDF]$. Using their coefficients to eliminate the terms with structure $FFFFFDDF$, one finds 7 total derivative terms with structure $FFFFDFDF$. Using them one can eliminate 7 terms in the contractions  $FFFFDFDF$. Using the Bianchi identity as we have done in the previous case, one finds 23 independent terms. 10 of them have $D_aF^{ab}$ which can be eliminated by appropriate field redefinitions. So up to field redefinitions there are the following 13 independent structures:
\begin{eqnarray}
  {S_p} \supset  - \alpha '{T_p}\int {{d^{p + 1}}\sigma \sqrt { - \det ({{\tilde G}_{ab}}  )} } \Big[&&\!\!\!\!\!\!\!\!\!\!\! {T_1}{F_{ae}}{F_{bf}}F_c^g{F_{dg}}{D^a}{F^{bc}}{D^d}{F^{ef}} \hfill \nonumber\\
   + {T_2}{F_{ab}}F_c^g{F_{dg}}{F_{ef}}{D^a}{F^{bc}}{D^d}{F^{ef}} &+& {T_3}F_a^fF_c^g{F_{dg}}{F_{ef}}{D^a}{F^{bc}}{D^d}F_b^e \hfill \nonumber\\
   + {T_4}F_a^fF_c^g{F_{df}}{F_{eg}}{D^a}{F^{bc}}{D_b}{F^{de}} &+& {T_5}F_a^f{F_{cf}}F_d^g{F_{eg}}{D^a}{F^{bc}}{D^d}F_b^e \hfill \nonumber\\
   + {T_6}{F_{ae}}{F_{bd}}F_c^g{F_{fg}}{D^a}{F^{bc}}{D^d}{F^{ef}} &+& {T_7}F_a^fF_c^g{F_{de}}{F_{fg}}{D^a}{F^{bc}}{D_b}{F^{de}} \hfill \nonumber\\
   + {T_8}{F_{ab}}{F_{ce}}F_d^g{F_{fg}}{D^a}{F^{bc}}{D^d}{F^{ef}} &+& {T_9}F_a^f{F_{ce}}F_d^g{F_{fg}}{D^a}{F^{bc}}{D^d}F_b^e \hfill \nonumber\\
   + {T_{10}}{F_{ad}}F_c^fF_e^g{F_{fg}}{D^a}{F^{bc}}{D_b}{F^{de}} &+& {T_{11}}{F_{ae}}{F_{cd}}{F_{fg}}{F^{fg}}{D^a}{F^{bc}}{D^d}F_b^e \hfill \nonumber\\
   + {T_{12}}{F_{ad}}{F_{ce}}{F_{fg}}{F^{fg}}{D^a}{F^{bc}}{D_b}{F^{de}}&+& {T_{13}}{F_{ac}}{F_{de}}{F_{fg}}{F^{fg}}{D^a}{F^{bc}}{D_b}{F^{de}}\Big] \labell{a17}
\end{eqnarray} 
The coefficients $T_1,\cdots, T_{13}$ are 13 parameters that we are going to find them by the T-duality constraint. 
 
There are 12 independent terms with structure $FFFF\Omega\Omega$. The terms that have  trace of the second fundamental form   may be eliminated by appropriate field redefinitions. The remaining terms are
\begin{eqnarray}
  {S_p} \supset  -{\alpha'} {T_p}\int {{d^{p + 1}}\sigma \sqrt { - \det ({{\tilde G}_{ab}} )} } \Big[ &&\!\!\!\!\!\!\!\!\!\!\! {W_1}F_a^{\,b}{F^{af}}F_c^{\,e}{F^{cd}}\Omega _{fd}^{\,\,\,\,\,\,\,i}{\Omega _{bei}} \hfill \nonumber \\
   + {W_2}F_a^{\,b}{F^{af}}F_f^{\,c}{F^{de}}\Omega _{bd}^{\,\,\,\,\,\,\,i}{\Omega _{cei}} &+& {W_3}{F_{af}}{F^{af}}{F^{bc}}{F^{de}}\Omega _{bd}^{\,\,\,\,\,\,\,i}{\Omega _{cei}} \hfill \nonumber\\
   + {W_4}F_a^{\,b}{F^{af}}F_c^{\,e}{F^{cd}}\Omega _{fb}^{\,\,\,\,\,\,\,i}{\Omega _{dei}} &+& {W_5}F_a^{\,b}{F^{af}}F_f^{\,c}F_b^{\,d}\Omega _c^{\,\,e\,i}{\Omega _{dei}} \hfill \nonumber\\
   + {W_6}{F_{af}}{F^{af}}F_b^{\,d}{F^{bc}}\Omega _c^{\,\,e\,i}{\Omega _{dei}} &+& {W_7}F_a^{\,b}{F^{af}}F_f^{\,c}{F_{bc}}{\Omega _{dei}}{\Omega ^{dei}} \hfill \nonumber\\
   + {W_8}{F_{af}}{F^{af}}{F_{bc}}{F^{bc}}{\Omega _{dei}}{\Omega ^{dei}} \Big] \labell{a15}
\end{eqnarray}
The coefficients $W_1,\cdots, W_8$ are 8 parameters that we are going to find them by the T-duality constraint. The parameters $T_1,\cdots, T_{13}$ and  $W_1,\cdots, W_8$ may also be found from studying the S-matrix element of six open string vertex operators. However, as far as we know, because of the very lengthy calculations involved in the S-matrix elements, these coefficients have not been found in the literature.

At the level of six extra $F$'s, one finds the following 37 independent couplings for $DFDF$:
\begin{eqnarray}
  {S_p} \supset  - \alpha '{T_p}\int {{d^{p + 1}}\sigma \sqrt { - \det ({{\tilde G}_{ab}} )} } [ &&\!\!\!\!\!\!\!\!\! {Z_1}{F_{de}}{F^{de}}{F_{fg}}{F^{fg}}{F_{hu}}{F^{hu}}{D^a}{F^{bc}}{D_b}{F_{ac}} \hfill \nonumber\\
   + {Z_2}F_a^g{F_{bg}}F_c^hF_d^u{F_{eh}}{F_{fu}}{D^a}{F^{bc}}{D^d}{F^{ef}} &+& {Z_3}{F_{ae}}F_b^gF_c^hF_d^u{F_{fh}}{F_{gh}}{D^a}{F^{bc}}{D^d}{F^{ef}} \hfill \nonumber\\
   + {Z_4}F_a^g{F_{be}}F_c^hF_d^u{F_{fh}}{F_{gu}}{D^a}{F^{bc}}{D^d}{F^{ef}} &+& {Z_5}F_a^fF_c^gF_d^hF_e^u{F_{fh}}{F_{gu}}{D^a}{F^{bc}}{D^d}F_b^e \hfill \nonumber\\
   + {Z_6}F_b^fF_c^gF_d^hF_e^u{F_{fh}}{F_{gu}}{D^a}{F^{bc}}{D_a}{F^{de}} &+& {Z_7}{F_{ad}}F_b^gF_c^h{F_{eg}}F_f^u{F_{hu}}{D^a}{F^{bc}}{D^d}{F^{ef}} \hfill \nonumber\\
   + {Z_8}F_a^fF_c^g{F_{df}}F_e^hF_g^u{F_{hu}}{D^a}{F^{bc}}{D_b}{F^{de}} &+& {Z_9}F_a^fF_c^g{F_{df}}F_e^hF_g^u{F_{hu}}{D^a}{F^{bc}}{D^d}F_b^e \hfill \nonumber\\
   + {Z_{10}}F_a^f{F_{cf}}F_d^gF_e^hF_g^u{F_{hu}}{D^a}{F^{bc}}{D^d}F_b^e &+& {Z_{11}}{F_{ae}}{F_{bd}}F_c^gF_f^hF_g^u{F_{hu}}{D^a}{F^{bc}}{D^d}{F^{ef}} \hfill \nonumber\\
   + {Z_{12}}{F_{ad}}{F_{be}}F_c^gF_f^hF_g^u{F_{hu}}{D^a}{F^{bc}}{D^d}{F^{ef}} &+& {Z_{13}}{F_{ab}}F_c^g{F_{de}}F_f^hF_g^u{F_{hu}}{D^a}{F^{bc}}{D^d}{F^{ef}} \hfill \nonumber\\
   + {Z_{14}}{F_{ab}}{F_{ce}}F_d^gF_f^hF_g^u{F_{hu}}{D^a}{F^{bc}}{D^d}{F^{ef}} &+& {Z_{15}}{F_{ad}}F_c^fF_e^gF_f^hF_g^u{F_{hu}}{D^a}{F^{bc}}{D_b}{F^{de}} \hfill \nonumber\\
   + {Z_{16}}{F_{ad}}F_c^fF_e^gF_f^hF_g^u{F_{hu}}{D^a}{F^{bc}}{D^d}F_b^e &+& {Z_{17}}{F_{bd}}F_c^fF_e^gF_f^hF_g^u{F_{hu}}{D^a}{F^{bc}}{D_a}{F^{de}} \hfill \nonumber\\
   + {Z_{18}}F_c^eF_d^fF_e^gF_f^hF_g^u{F_{hu}}{D^a}{F^{bc}}{D_a}F_b^d &+& {Z_{19}}F_c^eF_d^fF_e^gF_f^hF_g^u{F_{hu}}{D^a}{F^{bc}}{D_b}F_a^d \hfill \nonumber\\
   + {Z_{20}}F_d^f{F^{de}}F_e^gF_f^hF_g^u{F_{hu}}{D^a}{F^{bc}}{D_b}{F_{ac}} &+& {Z_{21}}{F_{ad}}{F_{ce}}F_f^h{F^{fg}}F_g^u{F_{hu}}{D^a}{F^{bc}}{D^d}F_b^e \hfill \nonumber\\
   + {Z_{22}}{F_{bd}}{F_{ce}}F_f^h{F^{fg}}F_g^u{F_{hu}}{D^a}{F^{bc}}{D_a}{F^{de}} &+& {Z_{23}}F_c^e{F_{de}}F_f^h{F^{fg}}F_g^u{F_{hu}}{D^a}{F^{bc}}{D_a}F_b^d \hfill \nonumber\\
   + {Z_{24}}F_c^e{F_{de}}F_f^h{F^{fg}}F_g^u{F_{hu}}{D^a}{F^{bc}}{D_b}F_a^d &+& {Z_{25}}{F_{de}}{F^{de}}F_f^h{F^{fg}}F_g^u{F_{hu}}{D^a}{F^{bc}}{D_b}{F_{ac}} \hfill \nonumber\\
   + {Z_{26}}F_a^fF_c^g{F_{df}}{F_{eg}}{F_{hu}}{F^{hu}}{D^a}{F^{bc}}{D^d}F_b^e &+& {Z_{27}}F_b^fF_c^g{F_{df}}{F_{eg}}{F_{hu}}{F^{hu}}{D^a}{F^{bc}}{D_a}{F^{de}} \hfill \nonumber\\
   + {Z_{28}}F_a^f{F_{cf}}F_d^g{F_{eg}}{F_{hu}}{F^{hu}}{D^a}{F^{bc}}{D^d}F_b^e &+& {Z_{29}}{F_{ae}}{F_{bd}}F_c^g{F_{fg}}{F_{hu}}{F^{hu}}{D^a}{F^{bc}}{D^d}{F^{ef}} \hfill \nonumber\\
   + {Z_{30}}{F_{ad}}{F_{be}}F_c^g{F_{fg}}{F_{hu}}{F^{hu}}{D^a}{F^{bc}}{D^d}{F^{ef}} &+& {Z_{31}}{F_{ab}}F_c^g{F_{de}}{F_{fg}}{F_{hu}}{F^{hu}}{D^a}{F^{bc}}{D^d}{F^{ef}} \hfill \nonumber\\
   + {Z_{32}}F_a^fF_c^g{F_{de}}{F_{fg}}{F_{hu}}{F^{hu}}{D^a}{F^{bc}}{D_b}{F^{de}} &+& {Z_{33}}F_c^eF_d^fF_e^g{F_{fg}}{F_{hu}}{F^{hu}}{D^a}{F^{bc}}{D_a}F_b^d \hfill \nonumber\\
   + {Z_{34}}F_c^eF_d^fF_e^g{F_{fg}}{F_{hu}}{F^{hu}}{D^a}{F^{bc}}{D_b}F_a^d &+& {Z_{35}}{F_{ad}}{F_{ce}}{F_{fg}}{F^{fg}}{F_{hu}}{F^{hu}}{D^a}{F^{bc}}{D^d}F_b^e \hfill \nonumber\\
   + {Z_{36}}F_c^e{F_{de}}{F_{fg}}{F^{fg}}{F_{hu}}{F^{hu}}{D^a}{F^{bc}}{D_a}F_b^d &+& {Z_{37}}F_c^e{F_{de}}{F_{fg}}{F^{fg}}{F_{hu}}{F^{hu}}{D^a}{F^{bc}}{D_b}F_a^d]\labell{a61} 
\end{eqnarray}
And the following 16 couplings for $\Omega\Omega$:
\begin{eqnarray}
  {S_p} \supset  -  \alpha '{T_p}\int {{d^{p + 1}}\sigma \sqrt { - \det ({{\tilde G}_{ab}}  )} } [ &&\!\!\!\!\!\!\!{Y_1}F_a^b{F^{ac}}F_c^dF_e^f{F^{eg}}F_g^h\Omega _{bf}^{\,\,\,\,\,\,\,i}{\Omega _{dhi}} \hfill \nonumber\\
   + {Y_2}F_a^b{F^{ac}}F_c^dF_b^eF_f^g{F^{fh}}\Omega _{dh}^{\,\,\,\,\,\,\,i}{\Omega _{egi}} &+& {Y_3}{F_{ab}}{F^{ab}}F_c^d{F^{ce}}F_f^g{F^{fh}}\Omega _{eh}^{\,\,\,\,\,\,\,i}{\Omega _{dgi}} \hfill \nonumber\\
   + {Y_4}F_a^b{F^{ac}}F_c^dF_b^eF_d^f{F^{gh}}\Omega _{eg}^{\,\,\,\,\,\,\,i}{\Omega _{fhi}} &+& {Y_5}{F_{ab}}{F^{ab}}F_c^d{F^{ce}}F_e^f{F^{gh}}\Omega _{dg}^{\,\,\,\,\,\,\,i}{\Omega _{fhi}} \hfill \nonumber\\
   + {Y_6}F_a^b{F^{ac}}F_c^d{F_{bd}}{F^{ef}}{F^{gh}}\Omega _{eg}^{\,\,\,\,\,\,\,i}{\Omega _{fhi}} &+& {Y_7}{F_{ab}}{F^{ab}}{F_{cd}}{F^{cd}}{F^{ef}}{F^{gh}}\Omega _{eg}^{\,\,\,\,\,\,\,i}{\Omega _{fhi}} \hfill \nonumber\\
   + {Y_8}F_a^b{F^{ac}}F_c^dF_b^eF_f^g{F^{fh}}\Omega _{de}^{\,\,\,\,\,\,\,i}{\Omega _{hgi}} &+& {Y_9}{F_{ab}}{F^{ab}}F_c^d{F^{ce}}F_f^g{F^{fh}}\Omega _{ed}^{\,\,\,\,\,\,\,i}{\Omega _{hgi}} \hfill \nonumber\\
   + {Y_{10}}F_a^b{F^{ac}}F_c^dF_b^eF_d^fF_e^g\Omega _f^{\,\,\,\,hi}{\Omega _{ghi}} &+& {Y_{11}}{F_{ab}}{F^{ab}}F_c^d{F^{ce}}F_e^fF_d^g\Omega _f^{\,\,\,\,hi}{\Omega _{ghi}} \hfill \nonumber\\
   + {Y_{12}}F_a^b{F^{ac}}F_c^d{F_{bd}}F_e^f{F^{eg}}\Omega _g^{\,\,\,\,hi}{\Omega _{fhi}} &+& {Y_{13}}{F_{ab}}{F^{ab}}{F_{cd}}{F^{cd}}F_e^f{F^{eg}}\Omega _g^{\,\,\,\,hi}{\Omega _{fhi}} \hfill \nonumber\\
   + {Y_{14}}F_a^b{F^{ac}}F_c^dF_b^eF_d^f{F_{ef}}{\Omega _{ghi}}{\Omega ^{ghi}} &+& {Y_{15}}{F_{ab}}{F^{ab}}F_c^d{F^{ce}}F_e^f{F_{df}}{\Omega _{ghi}}{\Omega ^{ghi}} \hfill \nonumber\\
   + {Y_{16}}{F_{ab}}{F^{ab}}{F_{cd}}{F^{cd}}{F_{ef}}{F^{ef}}{\Omega _{ghi}}{\Omega ^{ghi}}] \labell{a51} 
\end{eqnarray}
The coefficients $Z_1,\cdots, Z_{37}$ and $Y_1,\cdots, Y_{16}$ are 53 parameters that we are going to find them by the T-duality constraint. This construction of  independent terms can be used to find higher order couplings in which we are not interested in this paper. We will show in the next section that almost all parameters in the above couplings can be fixed by the T-duality constraint except two of them.

\section{T-duality constraint}

The T-duality   relates the bosonic string theory campactified on a circle with radius  $ \rho  $ to the same theory compactified on another circle with radius $ \frac{{{\alpha'}}}{\rho } $. It   relates the tension of $ {D_p} $-brane   to the tension of $ {D_{p - 1}} $-brane or $ {D_{p + 1}} $-brane, depending on whether the original $ {D_p} $-brane is along or orthogonal to the circle, respectively. Assuming the world-volume couplings of the  $ {D_p} $-brane in the string frame are independent of $p$, we expect the T-duality also relates the world-volume effective action of  $ {D_p} $-brane   to the effective action of   $ {D_{p - 1}} $-brane or $ {D_{p + 1}} $-brane, \ie
\begin{eqnarray}
{S_{{D_p}}}\mathop  \to \limits^T  {S_{{D_{p \pm 1}}}} \label{a18} 
\end{eqnarray}
This action can be expanded at low energy, \ie 
\begin{eqnarray}
{S_{{D_p}}} = \sum\limits_{n = 0}^\infty  {{{({\alpha'})}^n}S_{{D_p}}^{(n)}}  \label{a19} 
\end{eqnarray}
At order $\alpha'^0$ the action is given by the DBI action \reef{a1}.  At order $\alpha'$, there are infinite number of couplings depending on the number of  extra $F$'s in $DFDF$ and $\Omega\Omega$ couplings. At zero extra $F$, it is given by \reef{a6}, at two extra $F$'s it is given by \reef{a11},   at four extra $F$'s it is given by  \reef{a17} and \reef{a15}, and so on. We are not interested in this paper in the couplings at order $\alpha'$ with eight and higher extra $F$'s, and on the couplings at higher orders of $\alpha'$.

    When the T-duality transformations act along the killing coordinate $y$, and the $y$-direction is a world-volume,  then the  transformations at the leading order of $\alpha'$ are:  
  \begin{eqnarray}
  {A^y}&   \stackrel{T^{(0)}}{\longrightarrow} & {\chi ^y}\nonumber \\
  {A^{\tilde a}}& \stackrel{T^{(0)}}{\longrightarrow}& {A^{\tilde a}},{\mkern 1mu} {\mkern 1mu} {\mkern 1mu} {\mkern 1mu} {\mkern 1mu} {\mkern 1mu} {\chi ^i} \stackrel{T^{(0)}}{\longrightarrow} {\chi ^i} \labell{a20}
\end{eqnarray} 
where $\ta$ is the world-volume index which does not include the $y$-direction.   These transformations are expected to  receive $\alpha'$ corrections. That is, the T-duality operator has an $ {{\alpha'}} $ expansion:
  \begin{eqnarray}
T = \sum\limits_{n = 0}^\infty  {{{({\alpha'})}^n}{T^{(n)}}} \label{a21}
\end{eqnarray} 
where $ {T^{(0)}} $ is the transformation \reef{a20}.

The invariance of the effective actions at order $ {({\alpha'})^0} $ then means that  
  \begin{eqnarray}
{S_{D_p}^{(0)}} \stackrel{T^{(0)}}{\longrightarrow} {S_{D_{p-1}}^{(0)}} \labell{a25}
\end{eqnarray} 
where ${S_{D_p}^{(0)}}$ is the reduction of D$_p$-brane action at order $\alpha'^0$ on the circle.  At order $ {{\alpha'}} $, the action has two terms, \ie $ S_{D_p} = {S_{D_p}^{(0)}} + {\alpha'}{S_{D_p}^{(1)}} $. The invariance then means
 \begin{eqnarray}
  {S_{D_p}^{(1)}}& \stackrel{T^{(0)}}{\longrightarrow}& {S_{D_{p-1}}^{(1)}} + \delta S^{(1)} \hfill \nonumber\\
  {S_{D_p}^{(0)}}& \stackrel{T^{(0)}+T^{(1)}}{\longrightarrow}& {S_{D_{p-1}}^{(0)}} +\delta S'^{(1)}+\cdots \labell{a26}
\end{eqnarray} 
where dots represent terms at higher orders of $\alpha'$. The above relation indicates that  the extra term   $ \delta S^{(1)}$ which is produced by applying the T-duality transformation \reef{a20} on the reduction of action $S_{D_p}^{(1)}$ on the circle, should be canceled by applying the T-duality transformations at order  $\alpha'$  on the reduction of the action ${S_{D_p}^{(0)}}$.   Since the transformations are on the actions, one   may add total derivative terms $J^{(1)}$ to make the cancellation  happens. That is why we call the $\alpha'$ order term in the second line of \reef{a26} to be $ \delta S'^{(1)}$, \ie $ \delta S'^{(1)}+ \delta S^{(1)}+J^{(1)}=0$.  Note that   $\delta S^{(1)}$ contain only terms which involve $\chi^y$, so the corrections to the T-duality transformations and the total derivative terms in $J^{(1)}$   should  include only  terms which contain $\chi^y$. Similar T-duality transformations exist for the  effective actions at the higher orders of $ {\alpha'} $.
	
	Since the T-duality transformations affect $A_a$ and $\chi^i$, it is convenient to expand the effective action,   the T-duality transformations and total derivative terms  at order $\alpha'^n$ in terms of powers of $F$ and $\prt\chi$ as well\footnote{Using the transformations \reef{a20}, one may find the T-duality transformations of the covariant objects $F$, $DF$, $\tG$ and $\Omega$. Then one may find the $\alpha'$ corrections to these objects by using the T-duality constraint. In this paper, however, we use perturbation to rewrite the covariant action in terms of $F$ and $\prt\chi$ and then use the T-duality transformations \reef{a20} and their corresponding $\alpha'$-corrections.}, \ie
	\beqa
	S_{D_p}^{(n)}&=&\sum_{m=0}^{\infty} S_{D_p}^{(m,n)}\nn\\
	T^{(n)}&=&\sum_{m=0}^{\infty}  T^{(m,n)} \nn\\
	J^{(n)}&=&\sum_{m=0}^{\infty}  J^{(m,n)}
	\eeqa
where $m$ is the  power  of $F, \,\prt F$, $\prt\chi$, $\prt\prt\chi$  in   $ S_{D_p}^{(m,n)}$ and  $ J^{(m,n)}$,  and it is the extra  power  of $F$ and $\prt\chi$ on the right hand side of the T-duality transformation  $ T^{(m,n)}$. For example, for $m=2$ the action at order $\alpha'^0$ is 
\beqa
S_{D_p}^{(2,0)}&=&-T_p\int d^{p+1}\sigma\Big[\frac{1}{4}F_{ab}F_{cd}\eta^{ac}\eta^{bd}+\frac{1}{2}\prt_a\chi^i\prt_b\chi^j\eta_{ij}\eta^{ab}\Big]\labell{b1}
\eeqa
and the T-duality transformation is $T^{(2,0)}=0$. In fact, $T^{(0,0)}$ is given by \reef{a20} and  $T^{(m,0)}=0$ for $m\neq 0$. The transformation  $ {T^{(m,1)}} $ is  
 \begin{eqnarray}
  {A^y}&\mathop  \to \limits^{{T^{(m,1)}}}& \alpha' (\delta {\chi ^y})^{(m,1)}\nonumber \\
  {A^{\tilde a}}&\mathop  \to \limits^{{T^{(m,1)}}}& \alpha' (\delta{A^{\tilde a}})^{(m,1)},{\mkern 1mu} {\mkern 1mu} {\mkern 1mu} {\mkern 1mu} {\mkern 1mu} {\mkern 1mu} {\chi ^i}\mathop  \to \limits^{{T^{(m,1)}}} \alpha' (\delta {\chi ^i})^{(m,1)} \labell{a201}
\end{eqnarray}
where $(\delta {\chi ^y})^{(m,1)},\, (\delta \chi^i)^{(m,1)},\,(\delta A^{\tilde a})^{(m,1)}$  are all contractions of  one $\prt\prt\chi^y$, $\prt\prt\chi^i$ or $\prt F$ and $m$ number of $F$, $\prt\chi^y$ or $\prt\chi^i$ with arbitrary parameters. Each term should have at least one $\chi^y$. We expect these parameters to be found by the T-duality constraint.

The invariance of the effective actions at order $ {({\alpha'})^0} $ then means   
  \begin{eqnarray}
{S_{D_p}^{(m,0)}} \stackrel{T^{(0,0)}}{\longrightarrow}  {S_{D_{p-1}}^{(m,0)}} \labell{a251}
\end{eqnarray} 
 for any number of  $m$.  Using the T-duality transformation \reef{a20}, one finds that the transformation \reef{a251} is satisfied for any number of $m$. That means the DBI action is covariant under the T-duality transformation \reef{a20}, as expected.

 The invariance at order $\alpha'$   means
 \begin{eqnarray}
  {S_{D_p}^{(m,1)}}&\stackrel{T^{(0,0)}}{\longrightarrow} & {S_{D_{p-1}}^{(m,1)}} + \delta S^{(m,1)} \hfill \nonumber\\
  {S_{D_p}^{(n,0)}}&\stackrel{T^{(0,0)}+T^{(m-n,1)}}{\longrightarrow}  & {S_{D_{p-1}}^{(n,0)}} + \delta S_n^{(m,1)} +\cdots \nn
\end{eqnarray} 
where $2\leq n \leq m-2$, and dots represent terms at higher orders of $\alpha'$. Adding total derivative terms at   order $J^{(m,1)}$, one finds the T-duality constraint 
\beqa
 \sum_{n=2}^{m-2}\delta S_n^{(m,1)}  + \delta S^{(m,1)}+J^{(m,1)}=0
\eeqa
There are similar constraints for the couplings at higher orders of $\alpha'$. 

The above constraint may be used at each level of $m$ to fix the parameters of independent couplings that we have found in the previous section. The simplest case is  the action at the level of $m=2$. Since we have chosen the   coefficient in \reef{a6} to make no correction to the propagator, $S_{D_p}^{(2,1)}$ is a total derivative term. Hence, the T-duality constraint does not fix the parameter $C$ in this action.  However, one expects it should be related to all other parameters  at orders $m>2$, because this coefficient appears in all couplings with $m\geq  2$.
 
\subsection{Two extra $F$'s}

At order $\alpha'$, and at the level of $m=4$, there are two contributions to the action $S_{D_p}^{(4,1)}$. One contribution is coming from \reef{a6} and the other one from \reef{a11}. The parameters  $C, C_1,C_2,C_3,N_1,N_2,N_3$ appear in $S_{D_p}^{(4,1)}$. Then one should reduce it on the circle along the $y$-direction and use the T-duality transformation \reef{a20}. Then one should compare the result with  $S_{D_{p-1}}^{(4,1)}$. One finds
\beqa
 {S_{D_p}^{(4,1)}}&\stackrel{T^{(0,0)}}{\longrightarrow}& {S_{D_{p-1}}^{(4,1)}} +\delta S^{(4,1)}\labell{b3}
\eeqa
where $\delta S^{(4,1)}$ contains some non-zero terms at the level of $m=4$ which includes all parameters $C, C_1,C_2,C_3,N_1,N_2,N_3$. They  can not be canceled even by total derivative terms. This indicates that the T-duality transformations \reef{a20}  at order $\alpha'^0$ must receive $\alpha'$ corrections if the parameters $C, C_1,C_2,C_3,N_1,N_2,N_3$ are non-zero. 

Since we have chosen the couplings \reef{a6} to have no corrections to the propagators, we expect the $\alpha'$-corrections to the T-duality transformations \reef{a20} have no linear term. This steams from the fact that the S-matrix elements in string theory which have standard propagators, satisfy the Ward identity corresponding to the T-duality \cite{Garousi:2017fbe}. In other words, the field theory with standard propagators, should have no $\alpha'$-correction to the T-duality transformations at the linear order, \ie 
\beqa
T^{(0,n)}&=&0\,\,\,; \,n>0
\eeqa
Hence, the corrections to the T-duality transformations \reef{a20} are at orders  $T^{(2,1)},\, T^{(4,1)},\, T^{(6,1)},\cdots$,  $T^{(2,2)},\, T^{(4,2)},\, T^{(6,2)},\cdots$, and so on.

Therefore, the extra terms in $\delta S^{(4,1)}$ should be canceled by the T-duality transformation   $T^{(2,1)}$ on the reduction of the action $S_{D_p}^{(2,0)}$ in \reef{b1}, \ie
\beqa
 {S_{D_p}^{(2,0)}}&\stackrel{T^{(0,0)}+T^{(2,1)}}{\longrightarrow} & {S_{D_{p-1}}^{(2,0)}} + \delta S_2^{(4,1)}+ \delta S_2^{(6,2)}\labell{b2}
\eeqa 
where  $\delta S_2^{(6,2)}$ contains some non-zero terms at order $\alpha'^2$ and at the level of $m=6$ in which we are not interested.  
The reduction of \reef{b1} is 
\begin{eqnarray}
{S_{D_p}^{(2,0)}} &=&  - {T_p}(2\pi\rho)\int d^p\sigma [\frac{1}{2}{\partial _{\tilde a}}{A_y}{\partial ^{\tilde a}}{A^y} + \frac{1}{2}{\partial _{\tilde a}}{\chi _i}{\partial ^{\tilde a}}{\chi ^i} - \frac{1}{2}{\partial _{\tilde a}}{A_{\tilde b}}{\partial ^{\tilde b}}{A^{\tilde a}} + \frac{1}{2}{\partial _{\tilde b}}{A_{\tilde a}}{\partial ^{\tilde b}}{A^{\tilde a}}] \labell{a31}
\end{eqnarray} 
 The T-duality transformation $T^{(2,1)}$ for  $(\delta\chi^y)^{(2,1)}$, $(\delta A^{\ta})^{(2,1)}$ , $(\delta\chi^i)^{(2,1)}$  are all contractions of the following expressions   by the flat metric $\eta^{\ta\tb}$ and with arbitrary coefficients:
 \begin{eqnarray}
 (\delta\chi^y)^{(2,1)} &\sim&  \partial _\ta \partial _\tb {\chi ^y}\partial_\tc {\chi ^y}\partial_\td {\chi ^y} + \partial_\ta \partial_\tb {\chi ^y}\partial_\tc \chi^i \partial_\td \chi^j\eta_{ij}  + \partial _\ta\partial_\tb \chi^i \partial_\tc \chi^j \partial_\td {\chi ^y}\eta_{ij} \nn\\
&&+ \partial_\ta \partial_\tb {\chi ^y}F_{\tc\td}F_{\te\tf} + \partial_\ta {\chi ^y}F_{\tb\tc}\partial_\td F_{\te\tf}\,, \nn\\
(\delta A^{\ta})^{(2,1)} &\sim& \partial_\ta {\chi ^y}\partial_\tb {\chi ^y}\partial_\tc F_{\td\te} + \partial_\ta \partial_\tb {\chi ^y}\partial_\tc {\chi ^y}F_{\td\te}\,,\nn\\
(\delta\chi^i)^{(2,1)} &\sim& \partial_\ta \partial_\tb {\chi ^y}\partial_\tc {\chi ^y}\partial_\td \chi^i  + \partial_\ta \partial_\tb \chi^i \partial_\tc {\chi ^y}\partial_\td {\chi ^y}\,. \labell{a34}
\end{eqnarray}
Since the contractions involve derivatives of the field strength, one should impose the Bianchi identity $\prt_{[\ta}F_{\tb\tc]}=0$ to find independent terms. We impose this identity at the end after finding the parameters by the T-duality constraint. Applying  the above  T-duality transformations on \reef{b2}, one can find  $\delta S^{(4,1)}$ which contains the arbitrary parameters in \reef{a34}. To compare it with    $\delta S^{(4,1)}$ in \reef{b3}, one should also take into account the total derivative terms. 

The 
  total derivative terms can be written as
\beqa
J^{(4,1)}&=&-{T_{p-1}} \int {d^{p}}\sigma\eta^{\ta\tb}\prt_\ta {I}_\tb {}^{(4,1)}
\eeqa	
where  ${I}_\tb {}^{(4,1)}$ is all contractions with arbitrary parameters of the following expression with $\eta^{\ta\tb}$:
   \begin{eqnarray}
&& \partial_\ta \partial_\tb {\chi ^y}\partial_\tc {\chi ^y}\partial_\td {\chi ^y}\partial_\te {\chi ^y} + \partial_\ta \partial_\tb \chi^i \partial_\tc \chi^j \partial_\td {\chi ^y}\partial_\te {\chi ^y}\eta_{ij} + \partial_\ta \chi^i \partial_\tb \chi^j \partial_\tc \partial_\td {\chi ^y}\partial_\te {\chi ^y}\eta_{ij}\nn\\
&& + \partial_\ta \partial_\tb {\chi ^y}\partial_\tc {\chi ^y}F_{\td\te}F_{\te\tg} + \partial_\ta {\chi ^y}\partial_\tb {\chi ^y}F_{\tc\td}\partial_\te F_{\tf\tg} \label{a35}
\end{eqnarray}
Note that all   terms above and the terms in \reef{a34} involve $\chi^y$.

The T-duality constraint 
\beqa
\delta S^{(4,1)}+\delta S^{(4,1)}+J^{(4,1)}&=&0
\eeqa
Then gives some algebraic equations between the effective action parameters, the parameters in \reef{a34} and the parameters in the total derivative terms. On general ground, we do not expect the T-duality constraint fixes the overall coefficients of the T-dual multiplets. We choose $C=1$ which is fixed by the S-matrix calculation. Then if there is only one T-dual multiplet, its overall coefficient then should be fixed. The solution to the above equation produces the following relations between the effective action parameters  $ C_1,C_2,C_3,N_1,N_2,N_3$:
   \begin{eqnarray}
 {C_2} \to 1 &,&\,\,\,\,\,\,{C_1} \to 2 + 24{N_1}  ,\,\,\,\,\,\,{C_3} \to -\frac{1}{4}, \hfill \nonumber \\
  {N_3} \to -4N_1&,&\,\,\,\,\,\,N_2  \to  - 1 - 16{N_1}  \labell{a36}
\end{eqnarray}  
where the parameter $N_1$ remain arbitrary. This indicates that there are two T-dual multiplets, one multiplet with the overall coefficient $C=1$ and the second one with the overall coefficient $N_1$. As we will see, even though the parameter $N_1$ appears in the T-duality constraint at the levels $m>4$, the T-duality constraint at the levels of $m=6,8$ that we have done the calculations, can not fix this parameter.   The above parameters are consistent with the S-matrix calculation  results \reef{b4}, \ie if we choose the overall coefficient of the second multiplet to be $N_1=-1/24$, then the above parameters become exactly the S-matrix results in \reef{b4}. 
 
The algebraic equations at the level of $m=4$, gives the following $\alpha'$-corrections to the T-duality transformations:
\begin{eqnarray}
  {A^y} &\stackrel{ T^{(2,1)}}{\longrightarrow} & \alpha '[{E_1}{F^{\tilde b\tilde c}}{\partial _{\tilde a}}{F_{\tilde b\tilde c}}{\partial ^{\tilde a}}{\chi ^y} - (   1 + 12{N_1}){\partial _{\tilde a}}{\chi ^y}{\partial ^{\tilde a}}{\chi _y}{\partial _{\tilde b}}{\partial ^{\tilde b}}{\chi ^y} \hfill \nonumber\\
&&   + {E_3}{\partial _{\tilde a}}{\chi ^i}{\partial ^{\tilde a}}{\chi ^y}{\partial _{\tilde b}}{\partial ^{\tilde b}}{\chi _i} - (   1 +24{N_1}){\partial ^{\tilde a}}{\chi ^y}{\partial _{\tilde b}}{\partial _{\tilde a}}{\chi _y}{\partial ^{\tilde b}}{\chi ^y} \hfill \nonumber\\
&&   + 2{E_1}{\partial ^{\tilde a}}{\chi ^y}{F^{\tilde b\tilde c}}{\partial _{\tilde c}}{F_{\tilde a\tilde b}} + {E_2}{\partial ^{\tilde a}}{\chi ^y}F_{\tilde a}^{\tilde b}{\partial _{\tilde c}}F_{\tilde b}^{\tilde c} \hfill \nonumber\\
&&  - (   2 + 24{N_1})F_{\tilde a}^{\tilde c}{F^{\tilde a\tilde b}}{\partial _{\tilde c}}{\partial _{\tilde b}}{\chi ^y} + (\frac{1}{4} + 2{N_1}){\partial _{\tilde c}}{\partial ^{\tilde c}}{\chi ^y}{F_{\tilde a\tilde b}}{F^{\tilde a\tilde b}}] \hfill \nonumber\\
  {A^{\tilde a}} &\stackrel{ T^{(2,1)}}{\longrightarrow} & \alpha '[ - 4{N_1}{\partial _{\tilde b}}{\chi ^y}{\partial ^{\tilde b}}{\chi _y}{\partial _{\tilde c}}{F^{\tilde a\tilde c}} + (1 + 16{N_1}){\partial ^{\tilde a}}{\chi ^y}{\partial ^{\tilde b}}{\chi _y}{\partial _{\tilde c}}F_{\tilde b}^{\tilde c} \hfill \nonumber\\
&&   + (3 + 40{N_1}){\partial _{\tilde c}}{\partial ^{\tilde a}}{\chi _y}{\partial ^{\tilde b}}{\chi ^y}F_{\tilde b}^{\tilde c} + (2 + 32{N_1} - {E_2}){\partial _{\tilde c}}{\partial ^{\tilde c}}{\chi _y}{\partial ^{\tilde b}}{\chi ^y}F_{\tilde b}^{\tilde a} \hfill \nonumber\\
&&   + (1 + 24{N_1}){\partial ^{\tilde b}}{\chi ^y}{\partial ^{\tilde c}}{\chi _y}{\partial _{\tilde c}}F_{\tilde b}^{\tilde a}] \hfill \nonumber\\
  {\chi ^i}&\stackrel{ T^{(2,1)}}{\longrightarrow} & \alpha '[ - {E_3}{\partial _{\tilde a}}{\chi ^i}{\partial ^{\tilde a}}{\chi ^y}{\partial _{\tilde b}}{\partial ^{\tilde b}}{\chi _y} + {\partial ^{\tilde a}}{\chi ^y}{\partial _{\tilde b}}{\partial _{\tilde a}}{\chi ^i}{\partial ^{\tilde b}}{\chi _y} \hfill \nonumber\\
&&   - \frac{1}{2}{\partial _{\tilde a}}{\chi ^y}{\partial ^{\tilde a}}{\chi _y}{\partial _{\tilde b}}{\partial ^{\tilde b}}{\chi ^i}] \label{a42} 
\end{eqnarray}
where   $ {E_1} $, $ {E_2} $ and $ {E_3} $ are three other arbitrary parameters. However, the terms with coefficient $E_1$ cancels by using the Bianchi identity  $\prt_{[\ta}F_{\tb\tc]}=0$. So one can set $E_1=0$. The other two parameters may be fixed by studying the T-duality constraint at order $S^{(6,2)}$. Note that the above transformations are non-zero for any values for the parameters $E_2,E_3,N_1$. Hence, the T-duality constraint forces the leading order T-duality transformations \reef{a20} to receive higher derivative corrections.

If we have used the field redefinition freedom to remove the first term in \reef{a6}, the constraint \reef{a36} would not change, however, there would be a linear term $\prt\prt\chi^y$ in the T-duality transformation of $A^y$ and the coefficients of all terms in \reef{a42} would also change. The reason is that the T-duality transformations \reef{a42} are in fact the field redefinitions in the reduced space. The field redefinitions depends on whether or not we   keep the first term in \reef{a6}. 

\subsection{Four extra $F$'s}

At the order $\alpha'$ and at the level of $m=6$, there are three  contributions to the action $S_{D_p}^{(6,1)}$. One contribution is coming from \reef{a6}, another one is coming from \reef{a11} and the last one is coming from the couplings in \reef{a17} and \reef{a15}. The parameter  $ N_1 $ which has not been fixed in \reef{a36} and the parameters  $T_1,\cdots, T_{13}$ and $W_1,\cdots, W_8$ appear in $S_{D_p}^{(6,1)}$. One should reduce $S_{D_p}^{(6,1)}$ on the circle along the $y$-direction and use the T-duality transformation \reef{a20}. Then one should compare the result with  $S_{D_{p-1}}^{(6,1)}$. One finds
\beqa
 {S_{D_p}^{(6,1)}}&\stackrel{ T^{(0,0)}}{\longrightarrow} & {S_{D_{p-1}}^{(6,1)}} +\delta S^{(6,1)}\labell{b5}
\eeqa
where $\delta S^{(6,1)}$ contains some non-zero terms at the level of $m=6$ which includes all above parameters. Each term in $\delta S^{(6,1)}$ has the scalar field $\chi^y$.

The extra terms in $\delta S^{(6,1)}$ should be canceled by the T-duality transformation   $T^{(4,1)}$ on the reduction of the action $S_{D_p}^{(2,0)}$ in \reef{a31}, and by the T-duality transformation   $T^{(2,1)}$in \reef{a42}  on the reduction of the action $S_{D_p}^{(4,0)}$, \ie
\beqa
 {S_{D_p}^{(2,0)}}&\stackrel{ T^{(0,0)}+T^{(4,1)}}{\longrightarrow}  & {S_{D_{p-1}}^{(2,0)}} + \delta S_2^{(6,1)}+ \delta S_2^{(10,2)}\nn\\
{S_{D_p}^{(4,0)}}&\stackrel{ T^{(0,0)}+T^{(2,1)}}{\longrightarrow} & {S_{D_{p-1}}^{(4,0)}} + \delta S_4^{(6,1)}+ \delta S_4^{(8,2)} + \delta S_4^{(10,3)}+ \delta S_4^{(12,4)}\labell{b6}
\eeqa 
where  $\delta S_2^{(10,2)},\,\delta S_4^{(8,2)},\,  \delta S_4^{(10,3)}$ and $  \delta S_4^{(12,4)} $ contains some non-zero terms at higher orders of  $\alpha'$   in which we are not interested. 
It is straightforward to extract the action $S_{D_p}^{(4,0)}$    from  the DBI action \reef{a1} and then reduce it  on the circle along the $y$-direction. The T-duality transformation $T^{(2,1)}$ is given in \reef{a42}, and the T-duality transformation $T^{(4,1)}$ for  $(\delta\chi^y)^{(4,1)}$, $(\delta A^{\ta})^{(4,1)}$ , $(\delta\chi^i)^{(4,1)}$  are all contractions of the following expressions   by the flat metric $\eta^{\ta\tb}$ and with arbitrary coefficients:

\begin{eqnarray}
   (\delta\chi^y)^{(4,1)}&\sim&  \partial \partial {\chi ^y}\partial {\chi ^y}\partial {\chi ^y}\partial {\chi ^y}\partial {\chi ^y}+ \partial \partial {\chi ^y}\partial {\chi ^y}\partial {\chi ^y}\partial \chi \partial \chi    
   + \partial \partial \chi \partial \chi \partial {\chi ^y}\partial {\chi ^y}\partial {\chi ^y}+\partial \partial {\chi ^y}\partial \chi \partial \chi FF\nn\\
	&&+ \partial \partial {\chi ^y}\partial \chi \partial \chi \partial \chi \partial \chi  
   + \partial \partial \chi \partial \chi \partial \chi \partial \chi \partial {\chi ^y} + \partial \partial {\chi ^y}FFFF    
   + \partial {\chi ^y}FFF\partial F\nn\\   
  && + \partial \partial \chi \partial \chi \partial {\chi ^y}FF + \partial {\chi ^y}\partial \chi \partial \chi F\partial F    
   + \partial \partial {\chi ^y}\partial {\chi ^y}\partial {\chi ^y}FF+\partial {\chi ^y}\partial {\chi ^y}\partial {\chi ^y}F\partial F\,,  \nn\\
(\delta A^{\ta})^{(4,1)}	&\sim&\partial {\chi ^y}\partial {\chi ^y}FF\partial F + \partial \partial {\chi ^y}\partial {\chi ^y}FFF  
   + \partial {\chi ^y}\partial {\chi ^y}\partial \chi \partial \chi \partial F+\partial \partial {\chi ^y}\partial {\chi ^y}\partial \chi \partial \chi F\nn\\  
  && + \partial \partial \chi \partial \chi \partial {\chi ^y}\partial {\chi ^y}F + \partial {\chi ^y}\partial {\chi ^y}\partial {\chi ^y}\partial {\chi ^y}\partial F  
   + \partial \partial {\chi ^y}\partial {\chi ^y}\partial {\chi ^y}\partial {\chi ^y}F\,, \nn\\
(\delta\chi^i)^{(4,1)}	&\sim&\partial \partial \chi \partial {\chi^y}\partial {\chi^y}\partial {\chi^y}\partial {\chi^y} + \partial \partial {\chi^y}\partial {\chi^y}\partial {\chi^y}\partial {\chi^y}\partial \chi  
   + \partial \partial \chi \partial \chi \partial \chi \partial {\chi^y}\partial {\chi^y}\nn\\   
 &&  + \partial \partial {\chi^y}\partial {\chi^y}\partial \chi \partial \chi \partial \chi + \partial \partial \chi \partial {\chi^y}\partial {\chi^y}FF + \partial \partial {\chi^y}\partial {\chi^y}\partial \chi FF  
   + \partial {\chi^y}\partial {\chi^y}\partial \chi F\partial F \labell{a58}
\end{eqnarray}
where $ \partial  $ and $ F $ have $ (\tilde a,\tilde b,\tilde c,...) $ indices and $ \chi  $ has $ (i,j,k,...) $ indices.

We have to also consider total derivative terms, \ie 
\beqa
J^{(6,1)}&=&-{T_{p-1}} \int {d^{p}}\sigma\eta^{\ta\tb}\prt_\ta {I}_\tb {}^{(6,1)}
\eeqa	
where  ${I}_\tb {}^{(6,1)}$ is all contractions with arbitrary parameters of the following expression with $\eta^{\ta\tb}$:

\begin{eqnarray}
 &&  \partial \partial {\chi^y}\partial {\chi^y}\partial {\chi^y}\partial {\chi^y}\partial {\chi^y}\partial {\chi^y} + \partial \partial {\chi^y}\partial {\chi^y}\partial {\chi^y}\partial {\chi^y}\partial \chi \partial \chi   
   + \partial \partial \chi \partial \chi \partial {\chi^y}\partial {\chi^y}\partial {\chi^y}\partial {\chi^y}\nn\\
	&&+ \partial \partial \chi \partial \chi \partial \chi \partial \chi \partial {\chi^y}\partial {\chi^y}  
   + \partial \partial {\chi^y}\partial {\chi^y}\partial \chi \partial \chi \partial \chi \partial \chi +\partial \partial {\chi^y}\partial {\chi^y}\partial \chi \partial \chi FF \nn\\
	&& 
   + \partial \partial \chi \partial \chi \partial {\chi^y}\partial {\chi^y}FF+\partial {\chi^y}\partial {\chi^y}\partial \chi \partial \chi F\partial F  
   + \partial \partial {\chi^y}\partial {\chi^y}\partial {\chi^y}\partial {\chi^y}FF\nn\\
	&& + \partial {\chi^y}\partial {\chi^y}\partial {\chi^y}\partial {\chi^y}F\partial F  
   + \partial \partial {\chi^y}\partial {\chi^y}FFFF+ \partial {\chi^y}\partial {\chi^y}FFF\partial F  \labell{a59}
\end{eqnarray}
Note that all   terms above and the terms in \reef{a58} involve $\chi^y$.

Then the T-duality constraint 
\beqa
\delta S^{(6,1)}+\delta S_2^{(6,1)}+\delta S_4^{(6,1)}+J^{(6,1)}&=&0\labell{b7}
\eeqa
generates some algebraic equations between all  parameters.   The solution to these equations produce the following numbers for    the effective action parameters in \reef{a15} and \reef{a17}:
\begin{eqnarray}
&&  {W_1} \to 8{N_1},\,\,\,\,\,{W_2} \to  - 2 - 16{N_1},\,\,\,\,\,{W_3} \to \frac{1}{2} + 4{N_1},\,\,\,\,\,{W_4} \to  - 1 - 8{N_1}, \hfill \nonumber\\
&&  {W_5} \to 16{N_1},\,\,\,\,\,{W_6} \to \frac{1}{4} - 2{N_1},\,\,\,\,\,{W_7} \to \frac{1}{8},\,\,\,\,\,{W_8} \to  - \frac{1}{{32}}, \hfill \nonumber\\
&&  {T_1} \to 1,\,\,\,\,\,{T_2} \to 0,\,\,\,\,\,{T_3} \to  - \frac{2}{5} + \frac{{24}}{5}{N_1},\,\,\,\,\,{T_4} \to  - \frac{2}{5} + \frac{{24}}{5}{N_1}, \hfill \labell{a60}\\
&&  {T_5} \to \frac{7}{5} + \frac{{96}}{5}{N_1},\,\,\,\,\,{T_6} \to \frac{2}{5} - \frac{{24}}{5}{N_1},\,\,\,\,\,{T_7} \to  - \frac{1}{5} + \frac{{12}}{5}{N_1},\,\,\,\,\,{T_8} \to \frac{7}{5} - \frac{{24}}{5}{N_1}, \hfill \nonumber\\
&&  {T_9} \to  - \frac{3}{5} - \frac{{64}}{5}{N_1},\,\,\,\,\,{T_{10}} \to  - \frac{6}{5} - \frac{{48}}{5}{N_1},\,\,\,\,\,{T_{11}} \to 2{N_1},\,\,\,\,\,{T_{12}} \to \frac{1}{4} + 2{N_1},\,\,\,\,\,{T_{13}} \to {N_1}  \nn 
\end{eqnarray}
   The parameters in the first two lines  fix the action \reef{a15}. The other parameters fix the action \reef{a17}. The parameter $N_1$ could not be fixed by the calculation at the level $m=6$. So at this level there are two T-dual multiplets. However, from the S-matrix calculations in $m=4$ we know that $N_1=-1/24$.   It would be interesting to fix the parameters in \reef{a17}, \reef{a15} by the S-matrix calculations in $m=6$ and compare the result with the above numbers.

The parameters $E_2,E_3$ in the T-duality transformations $T^{(2,1)}$ appear in above calculations, however, the above T-duality constrain at the level $m=6$ could  not fix them. There are also many parameters in the  T-duality transformations $T^{(4,1)}$ which are not fix by the above calculations.  The  T-duality transformations  $T^{(4,1)}$ that   our calculation fixes appear in the appendix.

\subsection{Six extra $F$'s}

At the order $\alpha'$ and at the level of $m=8$, there are four  contributions to the action $S_{D_p}^{(8,1)}$. One contribution is coming from expanding \reef{a6} and keeping $m=8$ terms, the second contribution is coming from expanding \reef{a11} with the coefficients \reef{a36}, the third  contribution is coming   from expanding the couplings in \reef{a17} and \reef{a15} with the parameters \reef{a60}, and the last one is coming from the couplings in \reef{a61} and \reef{a51}. The parameter  $ N_1 $  and the parameters  $Z_1,\cdots, Z_{37}$ and $Y_1,\cdots, Y_{16}$ appear in $S_{D_p}^{(8,1)}$. One should reduce $S_{D_p}^{(8,1)}$ on the circle along the $y$-direction and use the T-duality transformation \reef{a20}. Then one should compare the result with  $S_{D_{p-1}}^{(8,1)}$. One finds
\beqa
 {S_{D_p}^{(8,1)}}&\stackrel{ T^{(0,0)}}{\longrightarrow} & {S_{D_{p-1}}^{(8,1)}} +\delta S^{(8,1)}\labell{b58}
\eeqa
where $\delta S^{(8,1)}$ contains some non-zero terms at the level of $m=8$ which includes all above parameters. Each term in $\delta S^{(8,1)}$ has the scalar field $\chi^y$.

The extra terms in $\delta S^{(8,1)}$ should be canceled by the T-duality transformation  $T^{(6,1)}$ on the reduction of the action $S_{D_p}^{(2,0)}$,  by the T-duality transformation $T^{(4,1)}$ on the reduction of the action $S_{D_p}^{(4,0)}$, and by the T-duality transformation   $T^{(2,1)}$   on the reduction of the action $S_{D_p}^{(6,0)}$, \ie
\beqa
 {S_{D_p}^{(2,0)}}&\stackrel{ T^{(0,0)}+T^{(6,1)}}{\longrightarrow}  & {S_{D_{p-1}}^{(2,0)}} + \delta S_2^{(8,1)}+ \delta S_2^{(16,2)}\nn\\
{S_{D_p}^{(4,0)}}&\stackrel{ T^{(0,0)}+T^{(4,1)}}{\longrightarrow} & {S_{D_{p-1}}^{(4,0)}} + \delta S_4^{(8,1)}+ \delta S_4^{(12,2)} + \delta S_4^{(16,3)}+ \delta S_4^{(20,4)}\labell{b8}\\
{S_{D_p}^{(6,0)}}&\stackrel{ T^{(0,0)}+T^{(2,1)}}{\longrightarrow} & {S_{D_{p-1}}^{(6,0)}} + \delta S_6^{(8,1)}+ \delta S_6^{(10,2)} + \delta S_6^{(12,3)}+ \delta S_6^{(14,4)} + \delta S_6^{(16,5)}+ \delta S_6^{(18,6)}\nn
\eeqa 
where  $\delta S_2^{(16,2)},\,\cdots,\delta S_6^{(18,6)}  $ contains some non-zero terms at higher orders of  $\alpha'$   in which we are not interested. 
  The T-duality transformation $T^{(2,1)}$ is given in \reef{a42},  the T-duality transformation $T^{(4,1)}$ is given in the appendix and $T^{(6,1)}$ can easily be constructed with some arbitrary parameters similar to \reef{a58}.   Similar to \reef{a59}, one can construct the total derivative terms $J^{(8,1)}$. 
	Then the T-duality constraint 
\beqa
\delta S^{(8,1)}+\delta S_2^{(8,1)}+\delta S_4^{(8,1)}+\delta S_6^{(8,1)}+J^{(8,1)}&=&0\labell{b78}
\eeqa
generates some algebraic equations between all unknown parameters in the T-duality transformation, the total derivative terms and the parameters in \reef{a61} and \reef{a51}.  

 The solution to  equation \reef{b78} produces the following numbers for    the effective action parameters in  \reef{a51}:
\begin{eqnarray}
&&  {Y_1} \to \frac{7}{5} + \frac{{56}}{5}{N_1},\,\,\,\,\,{Y_2} \to \frac{2}{5} - \frac{{24}}{5}{N_1},\,\,\,\,\,{Y_3} \to 0,\,\,\,\,\,{Y_4} \to \frac{{14}}{5} + \frac{{192}}{5}{N_1}, \hfill \nonumber\\
&&  {Y_5} \to  - \frac{3}{4} - 10{N_1},\,\,\,\,\,{Y_6} \to  - \frac{1}{4} - 3{N_1},\,\,\,\,\,{Y_7} \to \frac{1}{{16}} + \frac{3}{4}{N_1},\,\,\,\,\,{Y_8} \to \frac{3}{5} - \frac{{56}}{5}{N_1}, \hfill \nonumber\\
&&  {Y_9} \to 2{N_1},\,\,\,\,\,{Y_{10}} \to 1,\,\,\,\,\,{Y_{11}} \to  - \frac{1}{4},\,\,\,\,\,{Y_{12}} \to  - \frac{1}{8}, \hfill \nonumber\\
&&  {Y_{13}} \to \frac{1}{{32}},\,\,\,\,\,{Y_{14}} \to  - \frac{1}{{12}},\,\,\,\,\,{Y_{15}} \to \frac{1}{{32}},\,\,\,\,\,{Y_{16}} \to  - \frac{1}{{384}}, \hfill \labell{c81}
\eeqa
 And the following numbers for the effective action in \reef{a61}:
\beqa
 && {Z_1} \to \frac{{1 + 8{N_1}}}{{1920}},\,\,\,\,\,{Z_2} \to \frac{{ - 7 - 56{N_1}}}{5},\,\,\,\,\,{Z_3} \to \frac{{ - 4 - 12{N_1}}}{5},\,\,\,\,\,{Z_4} \to \frac{{ - 1 + 72{N_1}}}{5}, \hfill \nonumber\\
&&  {Z_5} \to 1,\,\,\,\,\,{Z_6} \to  - \frac{7}{{10}} + \frac{{72}}{5}{N_1},\,\,\,\,\,{Z_7} \to \frac{{23 - 96{N_1}}}{{35}},\,\,\,\,\,{Z_8} \to \frac{{ - 8 + 96{N_1}}}{5} \hfill \nonumber\\
&&  {Z_9} \to \frac{{4 - 48{N_1}}}{{35}},\,\,\,\,\,{Z_{10}} \to \frac{{ - 9 - 72{N_1}}}{5},\,\,\,\,\,{Z_{11}} \to \frac{{ - 7 + 144{N_1}}}{5},\,\,\,\,\,{Z_{12}} \to \frac{{36 - 552{N_1}}}{{35}}, \hfill \nonumber\\
&&  {Z_{13}} \to \frac{{1 + 88{N_1}}}{5},\,\,\,\,\,{Z_{14}} \to \frac{{ - 8 + 56{N_1}}}{5},\,\,\,\,\,{Z_{15}} \to \frac{{ - 2 - 176{N_1}}}{5},\,\,\,\,\,{Z_{16}} \to \frac{{17 + 976{N_1}}}{{35}}, \hfill \nonumber\\
 && {Z_{17}} \to \frac{{1 + 88{N_1}}}{5},\,\,\,\,\,{Z_{18}} \to \frac{{ - 3 + 96{N_1}}}{5},\,\,\,\,\,{Z_{19}} \to \frac{{ - 1 - 48{N_1}}}{5},\,\,\,\,\,{Z_{20}} \to  - \frac{7}{{60}} - \frac{{34}}{{15}}{N_1}, \hfill \nonumber\\
&&  {Z_{21}} \to  - \frac{3}{{20}} - \frac{{11}}{5}{N_1},\,\,\,\,\,{Z_{22}} \to \frac{3}{{80}} + \frac{9}{5}{N_1},\,\,\,\,\,{Z_{23}} \to  - \frac{1}{4},\,\,\,\,\,{Z_{24}} \to \frac{3}{{20}} + \frac{6}{5}{N_1}, \hfill \nonumber\\
&&  {Z_{25}} \to \frac{3}{{80}} + \frac{3}{{10}}{N_1},\,\,\,\,\,{Z_{26}} \to \frac{1}{{20}} + \frac{{15}}{5}{N_1},\,\,\,\,\,{Z_{27}} \to \frac{3}{{20}} + \frac{{11}}{5}{N_1},\,\,\,\,\,{Z_{28}} \to  - \frac{1}{{20}} - \frac{{22}}{5}{N_1}, \hfill \nonumber\\
&&  {Z_{29}} \to \frac{{1 + 8{N_1}}}{5},\,\,\,\,\,{Z_{30}} \to  - \frac{1}{{10}} - \frac{4}{5}{N_1},\,\,\,\,\,{Z_{31}} \to  - \frac{1}{{20}} + \frac{8}{5}{N_1},\,\,\,\,\,{Z_{32}} \to 3{N_1}, \hfill \nonumber\\
&&  {Z_{33}} \to \frac{3}{{10}} + \frac{{32}}{5}{N_1},\,\,\,\,\,{Z_{34}} \to  - 2{N_1},\,\,\,\,\,{Z_{35}} \to  - \frac{1}{{160}} + \frac{{{N_1}}}{5},\,\,\,\,\,{Z_{36}} \to \frac{{ - 1 - 3{N_1}}}{{20}}, \hfill \nonumber\\
&&  {Z_{37}} \to \frac{1}{{160}} + \frac{{{N_1}}}{{20}} \labell{c82} 
\end{eqnarray}
The solution to  the equation \reef{b78} produces also the T-duality transformation $T^{(6,1)}$ which is very lengthy expression and has many unfixed parameters. It is not illuminating, so we do not write it. It is interesting to note that the T-duality constraint could fix all parameters in the actions \reef{a61} and \reef{a51}. The parameter $N_1$ could not be fixed by the T-duality constraint even at the level of $m=8$. So the two T-dual multiplets remain independent at the level of $m=8$. It seems if one extends the above calculation to $m>8$, one would find only higher $F$-corrections to the two T-dual multiplets.  

\section{Discussion}

In this paper, we have found that the constraint that the covariant effective actions   must be invariant under the T-duality transformation \reef{a20} plus their appropriate higher derivative corrections, fixes the independent couplings in the effective actions at order $\alpha'$ up to two parameters, \ie \reef{a36}, \reef{a60}, \reef{c81} and \reef{c82}.  Hence, the T-duality constraint dictates that there are two T-dual multiplets. One with overall factor $C$ and the other one with the overall factor $N_1$. We have chosen the overall factor of the first multiplet to be $C=1$ which is dictated by the S-matrix calculations. The S-matrix also fixes the overall coefficient of the other T-dual multiplet to be $N_1=-1/24$. 

Another approach for imposing the T-duality constraint is that one considers non-covariant action and constrain it to be invariant under the standard T-duality \reef{a20} without $\alpha'$-corrections. Then one should use non-covariant field redefinitions and total derivative terms to convert the non-covariant action to the covariant form \cite{Garousi:2018qes}. This method has been used in \cite{Garousi:2018qes} to reproduce the known bulk effective action of the bosonic string theory at order $\alpha'$. We have used this method and found exactly the relations \reef{a36} at four-field level and \reef{a60} at six-field level. That is, we have written all contractions of $F,\prt F, \prt\chi, \prt\prt\chi$ at order   $\alpha'$ and at the level of $m=4$. Then we constrain  it to be invariant under the T-duality transformation \reef{a20}. The resulting action converted to \reef{a11} by appropriate non-covariant field redefinitions and total derivative terms provided that the relations \reef{a36} are satisfied. Similar calculation at the level of $m=6$ produces the coefficients in \reef{a60}. 

A specific non-covariant D-brane action at order $\alpha'$ in the bosonic string theory has been written in \cite{Garousi:2015qgr} which is invariant under T-duality transformations \reef{a20} and includes all powers of $F$. It includes $\prt F, \prt\prt\chi$ and some matrices that   contains all powers of $F$ and $\prt \chi$.  We have expanded that action at the level of $m=4$ and use non-covariant field redefinitions and total derivative terms to convert it to the covariant action \reef{a11}. We have succeeded at the level of $m=4$, however, we could not found convariant action at the level of $m=6$. That means the action proposed in \cite{Garousi:2015qgr} does not produce the result of the S-matrix calculations at the level of $m>4$. In fact the $F$ and $\prt\chi$ in the matrices used in  \cite{Garousi:2015qgr}  must be constant. The same matrices have been used in \cite{Ghodsi:2016qey} to construct the effective action of two massless closed strings and infinite number of constant $F$. It has been shown in \cite{Ghodsi:2016qey} that the result is consistent with the S-matrix element of two closed string vertex operators in the presence of constant $F$. 

We have found the couplings at order $\alpha'$ with zero, two, four and six extra $F$. In general there are non-zero couplings with more than six extra $F$ as well. One may try to find a closed expression for all couplings at order $\alpha'$. One suggestion may be to extend the pull-back metric $\tG^{ab}$ in \reef{a6} to include $F$'s as well. An extension is the following symmetric matrix: 
\begin{eqnarray}
{G^{ab}} = {\left( {\frac{1}{{\tG + F}}\tG\frac{1}{{\tG - F}}} \right)^{ab}} \label{a78}
\end{eqnarray}
In the absence of the transverse scalar fields $\chi^i$, it is the open string metric   which appears in the effective action when it is written in terms  of  non-commutative variables \cite{Seiberg:1999vs}. In terms of the commutative variables which we are working with, the above matrix may be used to rewrite the couplings we have found by the T-duality constraint in a closed expression. For example, all the couplings which have $\Omega_{ab}{}^i\Omega^{ab}{}_i$ can be written as
\begin{eqnarray}
{S_p} &\supset&    {\alpha'}{T_p}\int {{d^{p + 1}}\sigma \sqrt { - \det ({{  G}_{ab}} )} } \Big[  \Omega_{ab}{}^i\Omega^{ab}{}_i \Big]
 \end{eqnarray}
where $\det (  G_{ab} )=\det(\tilde G_{ab}+F_{ab})$. Expanding the DBI part, it produces all couplings we have found in \reef{a36}, \reef{a60}, \reef{c81} and \reef{c82} which includes the  structure $\Omega_{ab}{}^i\Omega^{ab}{}_i$. To be able to rewrite all other  couplings in a closed expression, one may   also need the following antisymmetric matrix as well:
\begin{eqnarray}
{\Theta^{ab}} = {\left( {\frac{1}{{\tG + F}}F\frac{1}{{\tG - F}}} \right)^{ab}} \label{a781}
\end{eqnarray}
It would be interesting to find a closed expression for the couplings that the T-duality constraint fixes. That expression would produce correct couplings with arbitrary number of $F$'s.

We have found the world-volume couplings at order $\alpha'$. One may be interested in extending these couplings to the order $\alpha'^2$. In this case, one should first find the independent couplings at order $\alpha'^2$ as we have done in section 2 for the couplings at order $\alpha'$. Then one should transform them under the T-duality transformation \reef{a20} at order $\alpha'^0$  to find $\delta S^{(m,2)}$. It should be canceled by total derivative terms $J^{(m,2)}$ and  by $\delta S_n^{(m,2)}$ terms which are resulted from transforming the DBI action under the T-duality transformations at order $\alpha'^2$ and from transforming the couplings at order $\alpha'$  under the T-duality transformations at order $\alpha'$ that we have found in this paper. This later terms makes the calculation in the bosonic theory to be very lengthy. However, in  the superstring theory there is no couplings at order $\alpha'$. Hence, the calculation would be much easier to perform. It would be interesting to find the $\alpha'^2$ corrections to the DBI and WZ actions in the superstring theory by the T-duality constraint and compare them with the couplings found in \cite{ Wyllard:2000qe, Wyllard:2001ye} by the boundary state formalism in superstring theory. 

{\bf Acknowledgments}:   This work is supported by Ferdowsi University of Mashhad under grant  3/44796(1396/08/02).
\newpage
{\Large{\bf Appendix}}

In this appendix, we write the T-duality transformation $T^{(4,1)}$ that the T-duality constraint \reef{b7} fixes. The transformation for $A^y$ is
\begin{eqnarray}
  {A^y} \stackrel{  T^{(4,1)}}{\longrightarrow} &\alpha '&\!\!\!\!\![ - 16{N_1}F_{\tilde a}^{\tilde c}{F^{\tilde a\tilde b}}F_{\tilde b}^{\tilde d}F_{\tilde c}^{\tilde e}{\partial _{\tilde d}}{\partial _{\tilde e}}{\chi ^y} + (\frac{1}{2} + 10{N_1}){F_{\tilde a\tilde b}}{F^{\tilde a\tilde b}}F_{\tilde c}^{\tilde e}{F^{\tilde c\tilde d}}{\partial _{\tilde d}}{\partial _{\tilde e}}{\chi ^y} \hfill \nonumber\\
   &-&\!\!\!\!\! (  \frac{{19}}{{10}} + \frac{{19}}{5}{N_1})F_{\tilde b}^{\tilde d}{F^{\tilde b\tilde c}}{\partial _{\tilde c}}{\partial _{\tilde d}}{\chi ^y}{\partial _{\tilde a}}{\chi _y}{\partial ^{\tilde a}}{\chi ^y} + (\frac{1}{4} + 4{N_1}){F_{\tilde b\tilde c}}{F^{\tilde b\tilde c}}{\partial ^{\tilde d}}{\partial _{\tilde d}}{\chi ^y}{\partial _{\tilde a}}{\chi _y}{\partial ^{\tilde a}}{\chi ^y} \hfill \nonumber\\
   &+&\!\!\!\!\! (2 + 24{N_1} - {E_3})F_{\tilde b}^{\tilde d}{F^{\tilde b\tilde c}}{\partial _{\tilde c}}{\partial _{\tilde d}}{\chi _i}{\partial _{\tilde a}}{\chi ^i}{\partial ^{\tilde a}}{\chi ^y} + {E_4}{F_{\tilde b\tilde c}}{F^{\tilde b\tilde c}}{\partial ^{\tilde d}}{\partial _{\tilde d}}{\chi _i}{\partial _{\tilde a}}{\chi ^i}{\partial ^{\tilde a}}{\chi ^y} \hfill \nonumber\\
   &+&\!\!\!\!\! (\frac{1}{4} + 8{N_1}){F_{\tilde c\tilde d}}{F^{\tilde c\tilde d}}{\partial _{\tilde a}}{\partial _{\tilde b}}{\chi ^y}{\partial ^{\tilde a}}{\chi ^y}{\partial ^{\tilde b}}{\chi _y} + \frac{2}{5}( - 1 + 12{N_1})F_{\tilde a}^{\tilde c}F_{\tilde c}^{\tilde d}{\partial _{\tilde b}}{\partial _{\tilde d}}{\chi ^y}{\partial ^{\tilde a}}{\chi _y}{\partial ^{\tilde b}}{\chi ^y} \hfill \nonumber\\
   &+&\!\!\!\!\! \frac{1}{5}(22 + 256{N_1} - 5{E_2})F_{\tilde a}^{\tilde c}F_{\tilde b}^{\tilde d}{\partial _{\tilde c}}{\partial _{\tilde d}}{\chi ^y}{\partial ^{\tilde a}}{\chi _y}{\partial ^{\tilde b}}{\chi ^y} \hfill \nonumber\\
   &-&\!\!\!\!\! (   \frac{1}{2} +6{N_1}){\partial ^{\tilde c}}{\partial _{\tilde c}}{\chi ^y}{\partial _{\tilde a}}{\chi _y}{\partial ^{\tilde a}}{\chi ^y}{\partial _{\tilde b}}{\chi ^y}{\partial ^{\tilde b}}{\chi _y} + (1 + 12{N_1} - {E_3}){\partial ^{\tilde c}}{\partial _{\tilde c}}{\chi ^y}{\partial _{\tilde a}}{\chi ^i}{\partial ^{\tilde a}}{\chi _y}{\partial _{\tilde b}}{\chi _i}{\partial ^{\tilde b}}{\chi ^y} \hfill \nonumber\\
   &+&\!\!\!\!\! {E_5}{\partial ^{\tilde c}}{\partial _{\tilde c}}{\chi _i}{\partial _{\tilde a}}{\chi ^y}{\partial ^{\tilde a}}{\chi _y}{\partial _{\tilde b}}{\chi ^i}{\partial ^{\tilde b}}{\chi ^y} - (  \frac{1}{4} + 2{N_1}){F_{\tilde c\tilde d}}{F^{\tilde c\tilde d}}{\partial _{\tilde a}}{\partial _{\tilde b}}{\chi ^y}{\partial ^{\tilde a}}{\chi ^i}{\partial ^{\tilde b}}{\chi _i} \hfill \nonumber\\
   &-&\!\!\!\!\! (   \frac{1}{2} + 4{N_1})F_{\tilde a}^{\tilde c}{F_{\tilde b\tilde c}}{\partial ^{\tilde d}}{\partial _{\tilde d}}{\chi ^y}{\partial ^{\tilde a}}{\chi ^i}{\partial ^{\tilde b}}{\chi _i} + {E_6}{\partial ^{\tilde c}}{\partial _{\tilde c}}{\chi _j}{\partial _{\tilde a}}{\chi ^i}{\partial ^{\tilde a}}{\chi ^y}{\partial _{\tilde b}}{\chi ^j}{\partial ^{\tilde b}}{\chi _i} \hfill \nonumber\\
   &-&\!\!\!\!\! {E_2}F_{\tilde a}^{\tilde c}F_{\tilde b}^{\tilde d}{\partial _{\tilde c}}{\partial _{\tilde d}}{\chi _i}{\partial ^{\tilde a}}{\chi ^y}{\partial ^{\tilde b}}{\chi ^i} + {E_7}F_{\tilde a}^{\tilde c}{F_{\tilde b\tilde c}}{\partial ^{\tilde d}}{\partial _{\tilde d}}{\chi _i}{\partial ^{\tilde a}}{\chi ^y}{\partial ^{\tilde b}}{\chi ^i} \hfill \nonumber\\
   &+&\!\!\!\!\! {E_8}{\partial ^{\tilde c}}{\partial _{\tilde c}}{\chi _i}{\partial _{\tilde a}}{\chi ^i}{\partial ^{\tilde a}}{\chi ^y}{\partial _{\tilde b}}{\chi _j}{\partial ^{\tilde b}}{\chi ^j} - 3(1 + 8{N_1}){\partial _{\tilde b}}{\partial _{\tilde c}}{\chi ^y}{\partial _{\tilde a}}{\chi _y}{\partial ^{\tilde a}}{\chi ^y}{\partial ^{\tilde b}}{\chi _y}{\partial ^{\tilde c}}{\chi ^y} \hfill \nonumber\\
   &+&\!\!\!\!\! (2 + 24{N_1} - {E_3}){\partial _{\tilde b}}{\partial _{\tilde c}}{\chi _i}{\partial _{\tilde a}}{\chi ^i}{\partial ^{\tilde a}}{\chi ^y}{\partial ^{\tilde b}}{\chi _y}{\partial ^{\tilde c}}{\chi ^y} - (   4 + 56{N_1} - {E_2})F_{\tilde a}^{\tilde d}{\partial ^{\tilde a}}{\chi ^y}{\partial ^{\tilde b}}{\chi _y}{\partial _{\tilde c}}{F_{\tilde b\tilde d}}{\partial ^{\tilde c}}{\chi ^y} \hfill \nonumber\\
   &+&\!\!\!\!\! (1 + 12{N_1}){\partial _{\tilde b}}{\partial _{\tilde c}}{\chi ^y}{\partial _{\tilde a}}{\chi _y}{\partial ^{\tilde a}}{\chi ^y}{\partial ^{\tilde b}}{\chi ^i}{\partial ^{\tilde c}}{\chi _i} + {E_2}F_{\tilde a}^{\tilde d}{\partial ^{\tilde a}}{\chi ^y}{\partial ^{\tilde b}}{\chi ^i}{\partial _{\tilde c}}{F_{\tilde b\tilde d}}{\partial ^{\tilde c}}{\chi _i} \hfill \nonumber\\
  & -&\!\!\!\!\! {E_3}{\partial _{\tilde b}}{\partial _{\tilde c}}{\chi _i}{\partial _{\tilde a}}{\chi ^i}{\partial ^{\tilde a}}{\chi ^y}{\partial ^{\tilde b}}{\chi ^j}{\partial ^{\tilde c}}{\chi _j} + \frac{1}{2}{F^{\tilde c\tilde d}}{\partial _{\tilde a}}{\chi ^y}{\partial ^{\tilde a}}{\chi _y}{\partial ^{\tilde b}}{\chi ^y}{\partial _{\tilde d}}{F_{\tilde b\tilde c}} \hfill \nonumber\\
   &+&\!\!\!\!\! {E_9}F_{\tilde b}^{\tilde c}{\partial _{\tilde a}}{\chi ^y}{\partial ^{\tilde a}}{\chi _y}{\partial ^{\tilde b}}{\chi ^y}{\partial _{\tilde d}}F_{\tilde c}^{\tilde d} + {E_{10}}F_{\tilde b}^{\tilde c}{\partial _{\tilde a}}{\chi ^i}{\partial ^{\tilde a}}{\chi ^y}{\partial ^{\tilde b}}{\chi _i}{\partial _{\tilde d}}F_{\tilde c}^{\tilde d} \hfill \nonumber\\
   &+&\!\!\!\!\! {E_{11}}F_{\tilde a}^{\tilde c}{\partial ^{\tilde a}}{\chi ^y}{\partial _{\tilde b}}{\chi _i}{\partial ^{\tilde b}}{\chi ^i}{\partial _{\tilde d}}F_{\tilde c}^{\tilde d} + {E_{12}}{F_{\tilde a\tilde b}}{\partial ^{\tilde a}}{\chi ^y}{\partial ^{\tilde b}}{\chi ^i}{\partial ^{\tilde c}}{\chi _i}{\partial _{\tilde d}}F_{\tilde c}^{\tilde d} \hfill \nonumber\\
   &+&\!\!\!\!\! \frac{3}{5}(1 + 8{N_1})F_{\tilde b}^{\tilde d}{F^{\tilde b\tilde c}}F_{\tilde c}^{\tilde e}{\partial ^{\tilde a}}{\chi ^y}{\partial _{\tilde e}}{F_{\tilde a\tilde d}} - \frac{1}{4}{F_{\tilde b\tilde c}}{F^{\tilde b\tilde c}}{F^{\tilde d\tilde e}}{\partial ^{\tilde a}}{\chi ^y}{\partial _{\tilde e}}{F_{\tilde a\tilde d}} \hfill \nonumber\\
   &+&\!\!\!\!\! \frac{1}{5}(17 + 256{N_1} - 5{E_2})F_{\tilde a}^{\tilde b}F_{\tilde c}^{\tilde e}{F^{\tilde c\tilde d}}{\partial ^{\tilde a}}{\chi ^y}{\partial _{\tilde e}}{F_{\tilde a\tilde d}} + {E_{13}}F_{\tilde a}^{\tilde b}{F_{\tilde c\tilde d}}{F^{\tilde c\tilde d}}{\partial ^{\tilde a}}{\chi ^y}{\partial _{\tilde e}}F_{\tilde b}^{\tilde e} \hfill \nonumber\\
   &-&\!\!\!\!\! F_{\tilde a}^{\tilde b}F_{\tilde b}^{\tilde c}{F^{\tilde d\tilde e}}{\partial ^{\tilde a}}{\chi ^y}{\partial _{\tilde e}}{F_{\tilde c\tilde d}} + {E_{14}}F_{\tilde a}^{\tilde b}F_{\tilde b}^{\tilde c}F_{\tilde c}^{\tilde d}{\partial ^{\tilde a}}{\chi ^y}{\partial _{\tilde e}}F_{\tilde d}^{\tilde e} \hfill \nonumber\\
   &+&\!\!\!\!\! (2 + 48{N_1}){\partial _{\tilde b}}{\partial _{\tilde c}}{\chi ^y}{\partial _{\tilde a}}{\chi ^i}{\partial ^{\tilde a}}{\chi _y}{\partial ^{\tilde b}}{\chi ^y}{\partial ^{\tilde c}}{\chi _i} + (2 + 24{N_1})F_{\tilde a}^{\tilde c}F_{\tilde b}^{\tilde d}{\partial _{\tilde c}}{\partial _{\tilde d}}{\chi ^y}{\partial ^{\tilde a}}{\chi ^i}{\partial ^{\tilde b}}{\chi _i} \hfill \nonumber\\
   &-&\!\!\!\!\! (   4 + 48{N_1})F_{\tilde a}^{\tilde c}F_{\tilde c}^{\tilde d}{\partial _{\tilde b}}{\partial _{\tilde d}}{\chi ^y}{\partial ^{\tilde a}}{\chi ^i}{\partial ^{\tilde b}}{\chi _i} - \frac{1}{8}F_{\tilde a}^{\tilde c}{F^{\tilde a\tilde b}}F_{\tilde b}^{\tilde d}{F_{\tilde c\tilde d}}{\partial ^{\tilde e}}{\partial _{\tilde e}}{\chi ^y} \hfill \nonumber\\
   &-&\!\!\!\!\! (  \frac{1}{{32}}+ \frac{{{N_1}}}{2}){F_{\tilde a\tilde b}}{F^{\tilde a\tilde b}}{F_{\tilde c\tilde d}}{F^{\tilde c\tilde d}}{\partial ^{\tilde e}}{\partial _{\tilde e}}{\chi ^y}- (  \frac{5}{2} + 32{N_1} - {E_2})F_{\tilde a}^{\tilde c}{F_{\tilde b\tilde c}}{\partial ^{\tilde d}}{\partial _{\tilde d}}{\chi ^y}{\partial ^{\tilde a}}{\chi _y}{\partial ^{\tilde b}}{\chi ^y} ]  \label{a9} 
\end{eqnarray}
where $E_2,E_3$ are the parameters that appear also in \reef{a42} which could  not be fixed by the constraint \reef{b7}. The other parameters $E_4, \cdots, E_{20}$  in above and the following transformations are also the free parameters that  the constraint \reef{b7} could not fixed. They may be fixed by considering the higher order constraints. The transformation for $A^\ta$ is
\begin{eqnarray}
  {A^{\tilde a}}  \stackrel{  T^{(4,1)}}{\longrightarrow} &\alpha '&\!\!\!\!\![-(  \frac{1}{4} + 10{N_1})F_{\tilde b}^{\tilde c}{F_{\tilde d\tilde e}}{F^{\tilde d\tilde e}}{\partial ^{\tilde a}}{\partial _{\tilde c}}{\chi ^y}{\partial ^{\tilde b}}{\chi _y} - \frac{1}{5}(   1 + 88{N_1})F_{\tilde b}^{\tilde c}F_{\tilde c}^{\tilde d}F_{\tilde d}^{\tilde e}{\partial ^{\tilde a}}{\partial _{\tilde e}}{\chi ^y}{\partial ^{\tilde b}}{\chi _y} \hfill \nonumber\\
   &-&\!\!\!\!\! \frac{1}{4}{F^{\tilde a\tilde c}}{F_{\tilde d\tilde e}}{F^{\tilde d\tilde e}}{\partial _{\tilde b}}{\partial _{\tilde c}}{\chi ^y}{\partial ^{\tilde b}}{\chi _y}+ (4 + 64{N_1} + {E_{14}} - {E_2}){F^{\tilde a\tilde c}}F_{\tilde b}^{\tilde d}{F_{\tilde c\tilde d}}{\partial ^{\tilde e}}{\partial _{\tilde e}}{\chi ^y}{\partial ^{\tilde b}}{\chi _y}  \hfill \nonumber\\
  & -&\!\!\!\!\! \frac{3}{5}(1 + 8{N_1}){F^{\tilde a\tilde c}}F_{\tilde b}^{\tilde d}F_{\tilde d}^{\tilde e}{\partial _{\tilde c}}{\partial _{\tilde e}}{\chi ^y}{\partial ^{\tilde b}}{\chi _y}- \frac{1}{5}(  17 + 256{N_1}){F^{\tilde a\tilde c}}F_{\tilde b}^{\tilde d}F_{\tilde c}^{\tilde e}{\partial _{\tilde d}}{\partial _{\tilde e}}{\chi ^y}{\partial ^{\tilde b}}{\chi _y} \hfill \nonumber\\
   &+&\!\!\!\!\! (\frac{{17}}{5} + \frac{{216}}{5}{N_1} + {E_2})F_{\tilde b}^{\tilde a}F_{\tilde c}^{\tilde e}{F^{\tilde c\tilde d}}{\partial _{\tilde d}}{\partial _{\tilde e}}{\chi ^y}{\partial ^{\tilde b}}{\chi _y} - {F^{\tilde a\tilde c}}F_{\tilde c}^{\tilde d}F_{\tilde d}^{\tilde e}{\partial _{\tilde b}}{\partial _{\tilde e}}{\chi ^y}{\partial ^{\tilde b}}{\chi ^y} \hfill \nonumber\\
   &-& \!\!\!\!\!(  \frac{3}{4} + 14{N_1} + {E_{13}})F_{\tilde b}^{\tilde a}{F_{\tilde c\tilde d}}{F^{\tilde c\tilde d}}{\partial ^{\tilde e}}{\partial _{\tilde e}}{\chi ^y}{\partial ^{\tilde b}}{\chi _y} + \frac{1}{5}(17 + 176{N_1})F_{\tilde b}^{\tilde d}{\partial _{\tilde c}}{\partial _{\tilde d}}{\chi ^y}{\partial ^{\tilde a}}{\chi _y}{\partial ^{\tilde b}}{\chi ^y}{\partial ^{\tilde c}}{\chi _y} \hfill \nonumber\\
   &-&\!\!\!\!\! \frac{1}{{10}}(   29 + 352{N_1})F_{\tilde c}^{\tilde d}{\partial ^{\tilde a}}{\partial _{\tilde d}}{\chi ^y}{\partial _{\tilde b}}{\chi _y}{\partial ^{\tilde b}}{\chi ^y}{\partial ^{\tilde c}}{\chi _y} + \frac{1}{{10}}(1 + 48{N_1}){F^{\tilde a\tilde d}}{\partial _{\tilde c}}{\partial _{\tilde d}}{\chi ^y}{\partial _{\tilde b}}{\chi _y}{\partial ^{\tilde b}}{\chi ^y}{\partial ^{\tilde c}}{\chi _y} \hfill \nonumber\\
   &-&\!\!\!\!\! (  8{N_1}+ {E_9} - {E_2})F_{\tilde c}^{\tilde a}{\partial ^{\tilde d}}{\partial _{\tilde d}}{\chi ^y}{\partial _{\tilde b}}{\chi _y}{\partial ^{\tilde b}}{\chi ^y}{\partial ^{\tilde c}}{\chi _y} - (   6 + 80{N_1} + {E_{15}})F_{\tilde c}^{\tilde d}{\partial ^{\tilde a}}{\partial _{\tilde d}}{\chi _i}{\partial _{\tilde b}}{\chi ^i}{\partial ^{\tilde b}}{\chi ^y}{\partial ^{\tilde c}}{\chi _y} \hfill \nonumber\\
   &-&\!\!\!\!\! (   2 + 32{N_1} + {E_{16}}- {E_2} + {E_3})F_{\tilde c}^{\tilde a}{\partial ^{\tilde d}}{\partial _{\tilde d}}{\chi _i}{\partial _{\tilde b}}{\chi ^i}{\partial ^{\tilde b}}{\chi ^y}{\partial ^{\tilde c}}{\chi _y} + {E_{15}}F_{\tilde c}^{\tilde d}{\partial ^{\tilde a}}{\partial _{\tilde d}}{\chi ^y}{\partial _{\tilde b}}{\chi ^i}{\partial ^{\tilde b}}{\chi _y}{\partial ^{\tilde c}}{\chi _i} \hfill \nonumber\\
   &-&\!\!\!\!\! \frac{1}{5}(  7 + 136{N_1})F_{\tilde b}^{\tilde d}F_{\tilde d}^{\tilde e}{\partial ^{\tilde b}}{\chi ^y}{\partial _{\tilde c}}F_{\tilde e}^{\tilde a}{\partial ^{\tilde c}}{\chi _y} + \frac{1}{5}(17 + 256{N_1}){F^{\tilde a\tilde d}}F_{\tilde d}^{\tilde e}{\partial ^{\tilde b}}{\chi ^y}{\partial _{\tilde c}}{F_{\tilde b\tilde e}}{\partial ^{\tilde c}}{\chi _y} \hfill \nonumber\\
   &+&\!\!\!\!\! ({E_{12}} + {E_2}){F_{\tilde b\tilde c}}{\partial ^{\tilde d}}{\partial _{\tilde d}}{\chi ^y}{\partial ^{\tilde a}}{\chi ^i}{\partial ^{\tilde b}}{\chi _y}{\partial ^{\tilde c}}{\chi _i}- (   \frac{1}{4} + 10{N_1}){F_{\tilde d\tilde e}}{F^{\tilde d\tilde e}}{\partial ^{\tilde b}}{\chi ^y}{\partial _{\tilde c}}F_{\tilde b}^{\tilde a}{\partial ^{\tilde c}}{\chi _y}  \hfill \nonumber\\
   &-& \!\!\!\!\!(  2 + 32{N_1}+ {E_{10}}- {E_3})F_{\tilde c}^{\tilde a}{\partial ^{\tilde d}}{\partial _{\tilde d}}{\chi ^y}{\partial _{\tilde b}}{\chi ^i}{\partial ^{\tilde b}}{\chi _y}{\partial ^{\tilde c}}{\chi _i} - (  1+ 16{N_1})F_{\tilde c}^{\tilde d}{\partial _{\tilde b}}{\partial _{\tilde d}}{\chi _i}{\partial ^{\tilde a}}{\chi ^y}{\partial ^{\tilde b}}{\chi _y}{\partial ^{\tilde c}}{\chi ^i} \hfill \nonumber\\
   &-& \!\!\!\!\!(   1 + 16{N_1} - {E_{17}}){F_{\tilde b\tilde c}}{\partial ^{\tilde d}}{\partial _{\tilde d}}{\chi _i}{\partial ^{\tilde a}}{\chi ^y}{\partial ^{\tilde b}}{\chi _y}{\partial ^{\tilde c}}{\chi ^i} + 4{N_1}F_{\tilde c}^{\tilde d}{\partial ^{\tilde a}}{\partial _{\tilde d}}{\chi _i}{\partial _{\tilde b}}{\chi ^y}{\partial ^{\tilde b}}{\chi _y}{\partial ^{\tilde c}}{\chi ^i} \hfill \nonumber\\
   &+&\!\!\!\!\! (\frac{1}{2} + 4{N_1} - {E_{18}})F_{\tilde c}^{\tilde a}{\partial ^{\tilde d}}{\partial _{\tilde d}}{\chi _i}{\partial _{\tilde b}}{\chi ^y}{\partial ^{\tilde b}}{\chi _y}{\partial ^{\tilde c}}{\chi ^i} - {E_{11}}F_{\tilde b}^{\tilde a}{\partial ^{\tilde d}}{\partial _{\tilde d}}{\chi ^y}{\partial ^{\tilde b}}{\chi _y}{\partial _{\tilde c}}{\chi _i}{\partial ^{\tilde c}}{\chi ^i} \hfill \nonumber\\
   &+&\!\!\!\!\! 4{N_1}{\partial _{\tilde b}}{\chi ^i}{\partial ^{\tilde b}}{\chi ^y}{\partial _{\tilde c}}{\chi _i}{\partial ^{\tilde c}}{\chi _y}{\partial _{\tilde d}}{F^{\tilde a\tilde d}} + (\frac{1}{2} + 4{N_1}){\partial ^{\tilde a}}{\chi ^y}{\partial _{\tilde b}}{\chi _y}{\partial ^{\tilde b}}{\chi ^y}{\partial ^{\tilde c}}{\chi _y}{\partial _{\tilde d}}F_{\tilde c}^{\tilde d} \hfill \nonumber\\
   &+&\!\!\!\!\! {E_{19}}{\partial ^{\tilde a}}{\chi ^i}{\partial _{\tilde b}}{\chi _i}{\partial ^{\tilde b}}{\chi ^y}{\partial ^{\tilde c}}{\chi _y}{\partial _{\tilde d}}F_{\tilde c}^{\tilde d} - (  1+ 16{N_1} - {E_{19}}){\partial ^{\tilde a}}{\chi ^y}{\partial _{\tilde b}}{\chi ^i}{\partial ^{\tilde b}}{\chi _y}{\partial ^{\tilde c}}{\chi _i}{\partial _{\tilde d}}F_{\tilde c}^{\tilde d} \hfill \nonumber\\
   &+&\!\!\!\!\! \frac{3}{5}(1 + {N_1}){F^{\tilde a\tilde d}}F_{\tilde b}^{\tilde e}{\partial ^{\tilde b}}{\chi ^y}{\partial ^{\tilde c}}{\chi _y}{\partial _{\tilde d}}{F_{\tilde c\tilde e}} + (\frac{{12}}{5} + \frac{{96}}{5}{N_1} + {E_2})F_{\tilde b}^{\tilde a}{\partial _{\tilde c}}{\partial _{\tilde d}}{\chi ^y}{\partial ^{\tilde b}}{\chi _y}{\partial ^{\tilde c}}{\chi ^y}{\partial ^{\tilde d}}{\chi _y} \hfill \nonumber\\
   &-&\!\!\!\!\! (   \frac{1}{2} + 16{N_1}){\partial _{\tilde b}}{\chi ^y}{\partial ^{\tilde b}}{\chi _y}{\partial ^{\tilde c}}{\chi ^y}{\partial _{\tilde d}}F_{\tilde c}^{\tilde a}{\partial ^{\tilde d}}{\chi _y} - (   4 + 64{N_1} + {E_{15}}){\partial _{\tilde b}}{\chi ^i}{\partial ^{\tilde b}}{\chi ^y}{\partial _{\tilde c}}F_{\tilde d}^{\tilde a}{\partial ^{\tilde c}}{\chi ^y}{\partial ^{\tilde d}}{\chi _i} \hfill \nonumber\\
   &-&\!\!\!\!\! (   6 + 80{N_1} + {E_{15}}){F_{\tilde b\tilde c}}{\partial ^{\tilde a}}{\partial _{\tilde d}}{\chi ^y}{\partial ^{\tilde b}}{\chi _y}{\partial ^{\tilde c}}{\chi ^i}{\partial ^{\tilde d}}{\chi _i} - (   2 + 32{N_1} - {E_2})F_{\tilde b}^{\tilde a}{\partial _{\tilde c}}{\partial _{\tilde d}}{\chi ^y}{\partial ^{\tilde b}}{\chi _y}{\partial ^{\tilde c}}{\chi ^i}{\partial ^{\tilde d}}{\chi _i} \hfill \nonumber\\
   &+&\!\!\!\!\! (2 + 16{N_1} + {E_{15}}){\partial _{\tilde b}}{\chi ^i}{\partial ^{\tilde b}}{\chi ^y}{\partial ^{\tilde c}}{\chi _y}{\partial _{\tilde d}}F_{\tilde c}^{\tilde a}{\partial ^{\tilde d}}{\chi _i} + 4{N_1}{\partial _{\tilde b}}{\chi ^y}{\partial ^{\tilde b}}{\chi _y}{\partial ^{\tilde c}}{\chi ^i}{\partial _{\tilde d}}F_{\tilde c}^{\tilde a}{\partial ^{\tilde d}}{\chi _i} \hfill \nonumber\\
   &-&\!\!\!\!\! (   1 + 16{N_1}){\partial ^{\tilde a}}{\chi ^y}{\partial ^{\tilde b}}{\chi _y}{\partial ^{\tilde c}}{\chi ^i}{\partial _{\tilde d}}{F_{\tilde b\tilde c}}{\partial ^{\tilde d}}{\chi _i} - (  2+16{N_1} + {E_{15}}){F_{\tilde b\tilde d}}{\partial ^{\tilde a}}{\partial _{\tilde c}}{\chi _i}{\partial ^{\tilde b}}{\chi ^y}{\partial ^{\tilde c}}{\chi _y}{\partial ^{\tilde d}}{\chi ^i} \hfill \nonumber\\
   &-&\!\!\!\!\! \frac{1}{5}(  3 + 104{N_1})F_{\tilde b}^{\tilde d}F_{\tilde d}^{\tilde e}{\partial ^{\tilde b}}{\chi ^y}{\partial ^{\tilde c}}{\chi _y}{\partial _{\tilde e}}F_{\tilde c}^{\tilde a} - (   \frac{1}{2} + 16{N_1})F_{\tilde c}^{\tilde e}{F^{\tilde c\tilde d}}{\partial _{\tilde b}}{\chi ^y}{\partial ^{\tilde b}}{\chi _y}{\partial _{\tilde e}}F_{\tilde d}^{\tilde a} \hfill \nonumber\\
   &-&\!\!\!\!\! 4{N_1}F_{\tilde b}^{\tilde d}{F_{\tilde c\tilde d}}{\partial ^{\tilde b}}{\chi ^y}{\partial ^{\tilde c}}{\chi _y}{\partial _{\tilde e}}{F^{\tilde a\tilde e}} + \frac{4}{5}(3 + 44{N_1})F_{\tilde c}^{\tilde e}{F^{\tilde c\tilde d}}{\partial ^{\tilde a}}{\chi ^y}{\partial ^{\tilde b}}{\chi _y}{\partial _{\tilde e}}{F_{\tilde b\tilde d}} \hfill \nonumber\\
   &-&\!\!\!\!\! (   \frac{1}{4} + 6{N_1}){F_{\tilde c\tilde d}}{F^{\tilde c\tilde d}}{\partial ^{\tilde a}}{\chi ^y}{\partial ^{\tilde b}}{\chi _y}{\partial _{\tilde e}}F_{\tilde b}^{\tilde e} - F_{\tilde b}^{\tilde c}{F^{\tilde d\tilde e}}{\partial ^{\tilde a}}{\chi ^y}{\partial ^{\tilde b}}{\chi _y}{\partial _{\tilde e}}{F_{\tilde c\tilde d}} \hfill \nonumber\\
   &+&\!\!\!\!\! \frac{1}{2}{F^{\tilde a\tilde c}}{F^{\tilde d\tilde e}}{\partial _{\tilde b}}{\chi ^y}{\partial ^{\tilde b}}{\chi _y}{\partial _{\tilde e}}{F_{\tilde c\tilde d}} - F_{\tilde b}^{\tilde a}{F^{\tilde d\tilde e}}{\partial ^{\tilde b}}{\chi ^y}{\partial ^{\tilde c}}{\chi _y}{\partial _{\tilde e}}{F_{\tilde c\tilde d}} \hfill \nonumber\\
   &+&\!\!\!\!\! {E_{20}}{F^{\tilde a\tilde d}}{F_{\tilde b\tilde d}}{\partial ^{\tilde b}}{\chi ^y}{\partial ^{\tilde c}}{\chi _y}{\partial _{\tilde e}}F_{\tilde c}^{\tilde e} - (   3 + 48{N_1} - {E_{20}})F_{\tilde b}^{\tilde c}F_{\tilde c}^{\tilde d}{\partial ^{\tilde a}}{\chi ^y}{\partial ^{\tilde b}}{\chi _y}{\partial _{\tilde e}}F_{\tilde d}^{\tilde e} \hfill \nonumber\\
   &+&\!\!\!\!\! (\frac{1}{2} + 12{N_1}){F^{\tilde a\tilde c}}F_{\tilde c}^{\tilde d}{\partial _{\tilde b}}{\chi ^y}{\partial ^{\tilde b}}{\chi _y}{\partial _{\tilde e}}F_{\tilde d}^{\tilde e} - (   1+ 16{N_1} + {E_2})F_{\tilde b}^{\tilde a}F_{\tilde c}^{\tilde d}{\partial ^{\tilde b}}{\chi ^y}{\partial ^{\tilde c}}{\chi _y}{\partial _{\tilde e}}F_{\tilde d}^{\tilde e} \hfill \nonumber\\
   &-&\!\!\!\!\! (   2 + 24{N_1})F_{\tilde d}^{\tilde a}{\partial _{\tilde b}}{\partial _{\tilde c}}{\chi _i}{\partial ^{\tilde b}}{\chi ^y}{\partial ^{\tilde c}}{\chi _y}{\partial ^{\tilde d}}{\chi ^i} + \frac{4}{5}(3 + 44{N_1})F_{\tilde b}^{\tilde d}F_{\tilde c}^{\tilde e}{\partial ^{\tilde b}}{\chi ^y}{\partial ^{\tilde c}}{\chi _y}{\partial _{\tilde e}}F_{\tilde d}^{\tilde a} \hfill \nonumber\\
   &+&\!\!\!\!\! 2{N_1}{\partial _{\tilde b}}{\chi ^y}{\partial ^{\tilde b}}{\chi _y}{\partial _{\tilde c}}{\chi ^y}{\partial ^{\tilde c}}{\chi _y}{\partial _{\tilde d}}{F^{\tilde a\tilde d}} + 2{N_1}{F_{\tilde c\tilde d}}{F^{\tilde c\tilde d}}{\partial _{\tilde b}}{\chi ^y}{\partial ^{\tilde b}}{\chi _y}{\partial _{\tilde e}}{F^{\tilde a\tilde e}}]  
\end{eqnarray}
The transformation for $\chi^i$ is
\begin{eqnarray}
  {\chi ^i} \stackrel{  T^{(4,1)}}{\longrightarrow} &\alpha '&\!\!\!\!\![-(   \frac{1}{2} + 4{N_1})F_{\tilde b}^{\tilde d}{F^{\tilde b\tilde c}}{\partial _{\tilde c}}{\partial _{\tilde d}}{\chi ^i}{\partial _{\tilde a}}{\chi ^y}{\partial ^{\tilde a}}{\chi _y} + {E_3}F_{\tilde b}^{\tilde d}{F^{\tilde b\tilde c}}{\partial _{\tilde c}}{\partial _{\tilde d}}{\chi ^y}{\partial _{\tilde a}}{\chi ^i}{\partial ^{\tilde a}}{\chi _y} \hfill \nonumber\\
   &-&\!\!\!\!\! (   \frac{1}{4} + 2{N_1} + {E_4}){F_{\tilde b\tilde c}}{F^{\tilde b\tilde c}}{\partial ^{\tilde d}}{\partial _{\tilde d}}{\chi ^y}{\partial _{\tilde a}}{\chi ^i}{\partial ^{\tilde a}}{\chi _y} - (   \frac{1}{2} + 2{N_1}){F_{\tilde c\tilde d}}{F^{\tilde c\tilde d}}{\partial _{\tilde a}}{\partial _{\tilde b}}{\chi ^i}{\partial ^{\tilde a}}{\chi ^y}{\partial ^{\tilde b}}{\chi _y} \hfill \nonumber\\
    &-& \!\!\!\!\!(   2+ 32{N_1})F_{\tilde a}^{\tilde c}F_{\tilde c}^{\tilde d}{\partial _{\tilde b}}{\partial _{\tilde d}}{\chi ^i}{\partial ^{\tilde a}}{\chi ^y}{\partial ^{\tilde b}}{\chi _y} + (1 + 16{N_1})F_{\tilde a}^{\tilde c}F_{\tilde b}^{\tilde d}{\partial _{\tilde c}}{\partial _{\tilde d}}{\chi ^i}{\partial ^{\tilde a}}{\chi ^y}{\partial ^{\tilde b}}{\chi _y} \hfill \nonumber\\
    &+&\!\!\!\!\! (\frac{1}{2} + 12{N_1} - {E_5} + {E_3}){\partial ^{\tilde c}}{\partial _{\tilde c}}{\chi ^y}{\partial _{\tilde a}}{\chi _y}{\partial ^{\tilde a}}{\chi ^y}{\partial _{\tilde b}}{\chi ^i}{\partial ^{\tilde b}}{\chi _y} + {E_3}{\partial ^{\tilde c}}{\partial _{\tilde c}}{\chi _j}{\partial _{\tilde a}}{\chi ^i}{\partial ^{\tilde a}}{\chi ^y}{\partial _{\tilde b}}{\chi ^j}{\partial ^{\tilde b}}{\chi _y} \hfill \nonumber\\
    &-&\!\!\!\!\! (   {E_7}- {E_2})F_{\tilde a}^{\tilde c}{F_{\tilde b\tilde c}}{\partial ^{\tilde d}}{\partial _{\tilde d}}{\chi ^y}{\partial ^{\tilde a}}{\chi _y}{\partial ^{\tilde b}}{\chi ^i} - (   {E_6} + {E_3}){\partial ^{\tilde c}}{\partial _{\tilde c}}{\chi ^y}{\partial _{\tilde a}}{\chi ^j}{\partial ^{\tilde a}}{\chi _y}{\partial _{\tilde b}}{\chi _j}{\partial ^{\tilde b}}{\chi ^i} \hfill \nonumber\\
   &-&\!\!\!\!\! {E_8}{\partial ^{\tilde c}}{\partial _{\tilde c}}{\chi ^y}{\partial _{\tilde a}}{\chi ^i}{\partial ^{\tilde a}}{\chi _y}{\partial _{\tilde b}}{\chi _j}{\partial ^{\tilde b}}{\chi ^j} + {E_3}{\partial _{\tilde b}}{\partial _{\tilde c}}{\chi ^y}{\partial _{\tilde a}}{\chi ^i}{\partial ^{\tilde a}}{\chi _y}{\partial ^{\tilde b}}{\chi ^y}{\partial ^{\tilde c}}{\chi _y} \hfill \nonumber\\
    &+&\!\!\!\!\! {E_3}{\partial _{\tilde b}}{\partial _{\tilde c}}{\chi ^y}{\partial _{\tilde a}}{\chi ^i}{\partial ^{\tilde a}}{\chi _y}{\partial ^{\tilde b}}{\chi ^j}{\partial ^{\tilde c}}{\chi _j} + {E_{17}}{F_{\tilde a\tilde c}}{\partial ^{\tilde a}}{\chi ^y}{\partial ^{\tilde b}}{\chi _y}{\partial ^{\tilde c}}{\chi ^i}{\partial _{\tilde d}}F_{\tilde b}^{\tilde d} \hfill \nonumber\\
    &+&\!\!\!\!\! {E_{16}}F_{\tilde b}^{\tilde c}{\partial _{\tilde a}}{\chi ^i}{\partial ^{\tilde a}}{\chi ^y}{\partial ^{\tilde b}}{\chi _y}{\partial _{\tilde d}}F_{\tilde c}^{\tilde d} - (   \frac{1}{2} - 12{N_1}){\partial _{\tilde b}}{\partial _{\tilde c}}{\chi ^i}{\partial _{\tilde a}}{\chi ^y}{\partial ^{\tilde a}}{\chi _y}{\partial ^{\tilde b}}{\chi ^y}{\partial ^{\tilde c}}{\chi _y} \hfill \nonumber\\
    &+&\!\!\!\!\! {E_{18}}F_{\tilde b}^{\tilde c}{\partial _{\tilde a}}{\chi ^y}{\partial ^{\tilde a}}{\chi _y}{\partial ^{\tilde b}}{\chi ^i}{\partial _{\tilde d}}F_{\tilde c}^{\tilde d} + \frac{1}{8}{F_{\tilde b\tilde c}}{F^{\tilde b\tilde c}}{\partial ^{\tilde d}}{\partial _{\tilde d}}{\chi ^i}{\partial _{\tilde a}}{\chi ^y}{\partial ^{\tilde a}}{\chi _y} \hfill \nonumber\\
    &+&\!\!\!\!\! \frac{3}{8}{\partial ^{\tilde c}}{\partial _{\tilde c}}{\chi ^i}{\partial _{\tilde a}}{\chi ^y}{\partial ^{\tilde a}}{\chi _y}{\partial _{\tilde b}}{\chi ^y}{\partial ^{\tilde b}}{\chi _y} + \frac{1}{2}{\partial ^{\tilde c}}{\partial _{\tilde c}}{\chi ^i}{\partial _{\tilde a}}{\chi ^j}{\partial ^{\tilde a}}{\chi ^y}{\partial _{\tilde b}}{\chi _j}{\partial ^{\tilde b}}{\chi _y} \hfill \nonumber\\
   &-&\!\!\!\!\! 2{\partial _{\tilde b}}{\partial _{\tilde c}}{\chi ^i}{\partial _{\tilde a}}{\chi ^j}{\partial ^{\tilde a}}{\chi ^y}{\partial ^{\tilde b}}{\chi _y}{\partial ^{\tilde c}}{\chi _j} + \frac{1}{2}{\partial _{\tilde b}}{\partial _{\tilde c}}{\chi ^i}{\partial _{\tilde a}}{\chi ^y}{\partial ^{\tilde a}}{\chi _y}{\partial ^{\tilde b}}{\chi ^j}{\partial ^{\tilde c}}{\chi _j}]  
\end{eqnarray}
In above transformations we have removed the terms that are canceled by the Bianchi identity $\prt_{[\ta}F_{\tb\tc]}=0$.
Note that the above transformations are non-zero for any specific values for the parameters $E_2,\cdots,E_{20}$. Hence, it is impossible to find solution for the T-duality constraint \reef{b7} without adding corrections to the standard T-duality transformations \reef{a20}.

\end{document}